\documentclass[journal]{IEEEtran}

\usepackage{booktabs}
\usepackage{amsmath,graphicx,cite,amssymb}
\usepackage{amsfonts,multirow,bm,array,setspace}

\usepackage{graphicx,float,cite,amssymb}
\usepackage{multirow,bm,array,setspace}
\usepackage{textcomp}
\usepackage{psfrag}
\usepackage{color}
\usepackage{pstricks}

\newcommand{\bX}{X}
\newcommand{\bY}{Y}

\newcommand{\bx}{\mathbf{x}}

\newcommand{\indep}{\mathrel{\text{\scalebox{1.07}{$\perp\mkern-10mu\perp$}}}}

\newcommand{\mV}{\mathsf{Var}}
\newcommand{\mC}{\mathsf{C}}
\newcommand{\bK}{\bm\Sigma}
\newcommand{\bG}{\mathbf G}
\newcommand{\bI}{\bm I}
\newcommand{\bQ}{\mathbf Q}

\newcommand{\bu}{\mathbf{u}}
\newcommand{\bv}{\mathbf{v}}
\newcommand{\bw}{\mathbf w}
\newcommand{\bPsi}{\bm\Lambda}
\newcommand{\Lmd}{\bm\Lambda}
\makeatletter
\newcommand{\distas}[1]{\mathbin{\overset{#1}{\kern\z@\sim}}}%
\newsavebox{\mybox}\newsavebox{\mysim}
\newcommand{\distras}[1]{%
  \savebox{\mybox}{\hbox{\kern1pt$\scriptstyle#1$\kern1pt}}%
  \savebox{\mysim}{\hbox{$\sim$}}%
  \mathbin{\overset{#1}{\kern\z@\resizebox{\wd\mybox}{\ht\mysim}{$\sim$}}}%
}
\makeatother
\ifCLASSINFOpdf
\else
\fi

\newtheorem{proposition}{Proposition}
\newtheorem{definition}{Definition}
\newtheorem{lemma}{Lemma}
\newtheorem{theorem}{Theorem}
\newtheorem{corollary}{Corollary}
\newtheorem{remark}{Remark}

\begin{document}
%
\title{Capacity Limits of Full-Duplex Cellular Network}

\author{\IEEEauthorblockN{Kaiming Shen, \IEEEmembership{Member,~IEEE}, Reza K. Farsani, \IEEEmembership{Student Member,~IEEE}, and Wei Yu, \IEEEmembership{Fellow,~IEEE}} 
\thanks{Manuscript received May 2, 2019; revised February 19, 2020; to appear in IEEE Transactions on Information Theory. The work of K. Shen was supported in part by National Natural Science Foundation of China (NSFC) 62001411, in part by Natural Sciences and Engineering Research Council (NSERC) of Canada, and in part by Huawei Technologies Canada. The work of R. K. Farsani and W. Yu was supported in part by NSERC of Canada, and in part by Huawei Technologies Canada. The materials in this paper have been presented in part in IEEE Information Theory Workshop (ITW),
December 2018, Guangzhou, China and in part in IEEE International Symposium on Information Theory (ISIT), July 2019, Paris, France. 

K. Shen is with the School of Science and Engineering, The Chinese University of Hong Kong (Shenzhen), Shenzhen 518172, China (e-mail: shenkaiming@cuhk.edu.cn).

R. K. Farsani and W. Yu are with
The Edward S. Rogers Sr. Department of Electrical and Computer
Engineering, University of Toronto, Toronto, ON M5S 3G4, Canada
(e-mails: \{rkfarsani, weiyu\}@ece.utoronto.ca).}%
}


%


\maketitle



\begin{abstract}
This paper aims to characterize the capacity limits of a wireless cellular
network with a full-duplex (FD) base-station (BS) and half-duplex user terminals,
in which three independent messages are communicated: the uplink message $m_1$
from the uplink user to the BS, the downlink message $m_2$ from the BS to the
downlink user, and the device-to-device (D2D) message $m_3$ from the uplink
user to the downlink user. From an information theoretical perspective, the
overall network can be viewed as a generalization of the FD relay broadcast
channel with a side message transmitted from the relay to the destination. We
begin with a simpler case that involves the uplink and downlink transmissions
of $(m_1,m_2)$ only, and propose an achievable rate region based on a novel
strategy that uses the BS as a FD relay to facilitate the interference
cancellation at the downlink user. We also prove a new converse, which is
strictly tighter than the cut-set bound, and characterize the capacity region
of the scalar Gaussian FD network without a D2D message to within a constant gap.
This paper further studies a general setup wherein $(m_1,m_2,m_3)$ are
communicated simultaneously. To account for the D2D message, we incorporate
Marton's broadcast coding into the previous scheme to obtain a larger
achievable rate region than the existing ones in the literature. We also
improve the cut-set bound by means of genie and show that by using one of the
two simple rate-splitting schemes, the capacity region of the scalar Gaussian
FD network with a D2D message can already be reached to within a constant gap.
Finally, a generalization to the vector Gaussian channel case is discussed.
Simulation results demonstrate the advantage of using the BS as relay in
enhancing the throughput of the FD cellular network.
\end{abstract}

\section{Introduction}

\IEEEPARstart{T}{raditional} wireless cellular systems separate uplink and
downlink signals by using either time division duplex (TDD) or frequency
division duplex (FDD), because at a conventional analog front-end, the echo
due to transmitting in one direction can overwhelm the receiver in the other
direction. Recent progress in analog and digital echo cancellation
\cite{choi2010achieving,aryafar2012midu,everett2014passive} is now opening
up the possibility of realizing bi-directional communication in a full-duplex
(FD) fashion. This paper considers the capacity limits of the FD cellular network.

\begin{figure}[t]
\begin{minipage}[b]{0.5\linewidth}
\psfrag{A}[][]{\small $m_1$}
\psfrag{C}[][]{\small $m_2$}
\psfrag{B}[][]{\small $m_3$}
\psfrag{1}[][]{\small 1}
\psfrag{2}[][]{\small 2}
\psfrag{3}[][]{\small 3}
  \centering
  \centerline{\includegraphics[width=3.5cm]{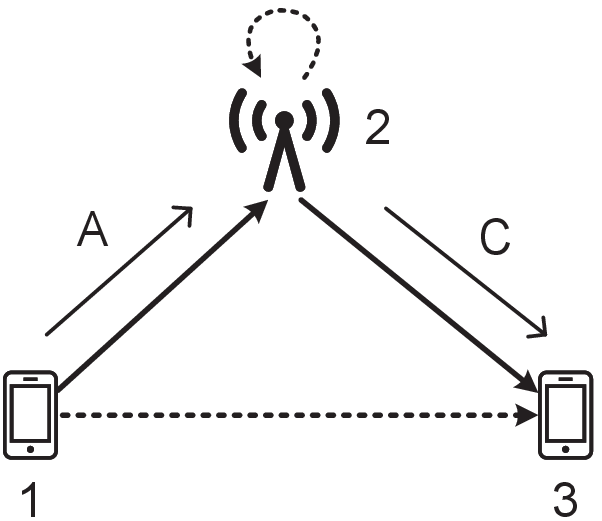}}
  \vspace{0.35em}
  \centerline{\footnotesize (a) Without a D2D message}
\end{minipage}
\hfill
\begin{minipage}[b]{0.48\linewidth}
\psfrag{A}[][]{\small $m_1$}
\psfrag{C}[][]{\small $m_2$}
\psfrag{B}[][]{\small $m_3$}
\psfrag{1}[][]{\small 1}
\psfrag{2}[][]{\small 2}
\psfrag{3}[][]{\small 3}
  \centering
  \centerline{\includegraphics[width=3.5cm]{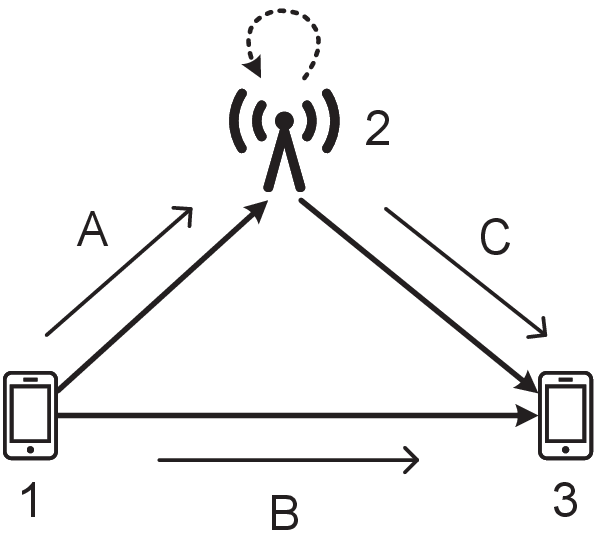}}
  \centerline{\footnotesize (b) With a D2D message}
\end{minipage}
	\caption{
Cellular network with uplink user (node 1), full-duplex BS (node 2),
and downlink user (node 3).  Here, $m_1$ is the uplink message, $m_2$ is the
downlink message, and $m_3$ is the D2D side-message. The loops represent
self-interference. The straight dashed line in (a) represents the
uplink-to-downlink cross-channel interference.}
\label{fig:model}
\end{figure}

In a FD cellular network, the base-station (BS) is capable of transmitting and
receiving signals in FD mode \cite{sab2014inband}; but it is often the case that
the uplink and downlink user terminals still operate in half-duplex mode.
In such a system as depicted in Fig.~\ref{fig:model}(a), although the uplink
transmission of $m_1$ and the downlink transmission of $m_2$ occupy the same
spectrum simultaneously thereby doubling the frequency-reuse factor as
compared to TDD or FDD, the cross-channel interference from the uplink user
(node 1) to the downlink user (node 3) is still a major source of
impairment. Such cross-channel interference is in fact the performance
bottleneck in FD networks as pointed out in \cite{xie2014double,GoyalCommMag2015},
especially when the uplink and downlink user terminals are in close proximity
to each other.

This paper aims to show that this cross-channel interference can potentially
be cancelled or significantly suppressed with the aid from the BS.  This is
because the BS can act as a relay, as it already needs to decode the uplink
message $m_1$, so it can help the downlink user cancel the cross-channel
interference due to $m_1$. Under a scalar Gaussian channel model for the setup
depicted in Fig.~\ref{fig:model}(a) with $(m_1,m_2)$ only, this paper shows
that the proposed interference cancellation scheme can achieve the capacity
region of this channel to within a constant gap.

\newcolumntype{C}[1]{>{\centering\arraybackslash}p{#1}}
\renewcommand{\arraystretch}{1.5}
\begin{table*}[t]
\renewcommand{\arraystretch}{1.3}
\small
\centering
\caption{\small Main Results of the Paper}
\begin{tabular}{|c||c|c|C{3.3cm}|}
\hline
& Achievability &  Converse & Capacity Region\\
\hline
\hline
Discrete Memoryless Channel (DMC) & Theorem \ref{theorem:DM:inner} & Theorem \ref{theorem:DM:outer} & --\\
\hline
Gaussian Channel & Proposition \ref{prop:G:inner} & Proposition \ref{prop:G:outer} & Theorems \ref{prop:G_very_strong:capacity}, \ref{theorem:G:capcity} and \ref{prop:G_strong:capacity}\\
\hline
\hline
DMC with D2D & Theorem \ref{theorem:DM_d2d:inner}, Corollaries \ref{theorem:DM_d2d:inner1} and \ref{theorem:DM_d2d:inner2} & Theorem \ref{theorem:DM_d2d:outer} & --\\
\hline
Gaussian Channel with D2D & Propositions \ref{prop:G_d2d:inner1} and \ref{prop:G_d2d:inner2} & Proposition \ref{prop:G_d2d:outer} and Corollary \ref{corollary:G_d2d:relaxed_OB} & Theorem \ref{theorem:G_d2d:gap}\\
\hline
\end{tabular}
\label{tab:main_results}
\end{table*}

This paper also considers a scenario in which in addition to the uplink and
downlink messages, the uplink user also wishes to directly send a separate
message $m_3$ to the downlink user via the device-to-device (D2D) link. For
this new channel model as shown in Fig.~\ref{fig:model}(b), 
we propose to incorporate Marton's broadcast coding \cite{MartonTIT1979}
of $m_1$ and $m_3$ into the previous transmission scheme to derive a general
achievable rate region. 
We further propose two simple rate-splitting schemes and a new converse, and
show that using one of the two rate-splitting schemes (depending on the channel
condition) already suffices to attain the capacity region to within a constant
gap for the scalar Gaussian FD cellular network with a D2D message.

{The FD cellular network with only the uplink and the downlink
transmissions has been extensively studied in the existing literature. Most
of the prior works propose to alleviate the cross-channel interference by
optimizing resource allocation, e.g., \cite{GoyalICC2014} schedules the uplink
and downlink users in accordance with the distance between them,
\cite{MarNET2017} uses power control to combat cross-channel interference and
self-interference, and \cite{YunTMC16,shen_vtc} further consider joint power
control and user scheduling. Moreover, \cite{ShaoCOML14} shows empirically that
the gain of FD mode over half-duplex increases with the number of users. For
the multiple-input multiple-output (MIMO) setup, \cite{KarakusISIT2015}
exploits spatial diversity by scheduling users in an opportunistic manner.
These optimization-based works always treat interference as noise. In contrast,
this paper employs more sophisticated coding techniques to try to cancel the
interference, while aiming to provide insight into the fundamental capacity
limits of the FD cellular network.}
In particular, the present work determines the capacity region to
within a constant gap in the scalar Gaussian channel case, as opposed to the existing
theoretical studies in \cite{chaeTWC18,chae2016limits,KhojastepourINFOCOM2015}
that only characterize the sum rate {in an asymptotic regime as
the signal-to-noise ratio (SNR) tends to infinity}. Furthermore, the capacity
analyses are extended to the FD cellular model with a D2D message.

We point out that the FD cellular network with D2D is equivalent to the relay
broadcast channel with side message (or with ``private'' message
\cite{Tannious07}). The authors of \cite{Tannious07} propose a
decode-and-forward scheme and a compress-and-forward scheme for this channel.
Our scheme is a further development of the decode-and-forward scheme
\cite{Tannious07} by incorporating multiple new techniques (including rate
splitting, joint decoding, and Marton's broadcast coding \cite{MartonTIT1979}).
With respect to the converse, \cite{Tannious07} derives an outer bound based on
the genie-aided method, but as indicated by the authors, the outer bound of
\cite{Tannious07} is not
computable. This paper develops better use of the auxiliary ``genie'' variables
to improve upon the cut-set bound, and further comes up with a new sum-rate
upper bound that plays a key role in characterizing the capacity region
for the scalar Gaussian case to within a constant gap.

The FD cellular network with D2D is also a generalization of the
\emph{partially-cooperative relay broadcast channel}
\cite{LiangTIT2007a,LiangTIT2007b} for which a modified Marton's broadcast
coding scheme has already been proposed. The achievability part of our paper
can be thought of as a generalization of
\cite{LiangTIT2007a,LiangTIT2007b} in incorporating the transmission of the
relay-to-destination side message $m_2$ into the modified Marton's coding.
Thus, the contribution of the present paper can also be thought of as the
characterization of the capacity of the Gaussian relay broadcast channel (with
side message) to within a constant gap.




For ease of reference, we categorize the main results of the paper in Table \ref{tab:main_results} as displayed at the top of the page. 
Specifically, the main contributions of this paper are as follows:
\begin{itemize}
    \item\emph{Achievability:} For the FD cellular network without a D2D message, we
propose a relaying strategy to improve upon the existing achievable
uplink and downlink rate region. When the D2D message is included, we
extend the scheme by incorporating Marton's broadcast coding. 
    
    \item\emph{Converse:} We derive new upper bounds on the sum rate for both
the cases with and without D2D. 
Further, we use different genies to provide tighter converses.
    
    \item\emph{Scalar Gaussian Channel:} 
We characterize the capacity region of the scalar Gaussian FD cellular network (both with
and without D2D) to within 1 bit in general. For the case without D2D, (i)
a smaller constant gap of approximately 0.6358 bits is established in the \emph{strong
interference} regime; (ii) the exact capacity is determined in the \emph{very
strong interference} regime.

    \item\emph{Vector Gaussian Channel:} We discuss the generalization of the achievability
results to the MIMO case that includes spatial multiplexing and dirty-paper coding.
\end{itemize}

\emph{Notation:} Let $[1:n]$ be the set $\{1,2,\ldots,n\}$, $\mathsf C(x)$ the function $\log_2(1+x)$ for $x\ge0$, $\mathbb R_+$ the set of nonnegative real numbers, and $\mathbb C$ the set of complex numbers. We use a superscripted letter to denote a sequence of variables, e.g.,
$X^N =(X_1,\ldots,X_N)$, use $\bm I$ to denote the identity matrix, and use $\mathbf A^H$ to denote the Hermitian transpose of matrix $\mathbf A$. For a random variable $X$, use $\mathbb E[X]$ to denote the expected value, and $\mathsf{Var}(X)$ the variance. For two random variables $X_1$ and $X_2$, use $\mathsf{Cov}(X_1,X_2)$ to denote their covariance, and use $X_1\indep X_2$ to indicate that they are independent.

The rest of the paper is organized as follows. Section \ref{sec:model} formally defines the various channel models. Section \ref{sec:DM} discusses the discrete memoryless channel. Section \ref{sec:scalar_G} discusses the scalar Gaussian channel. Section \ref{sec:vector_G} discusses the generalization to the vector Gaussian channel. Numerical results are presented in Section \ref{sec:simulation}. Finally, we conclude this work in Section \ref{sec:conclusion}.

\section{Full-Duplex Cellular Network Models}
\label{sec:model}

This work examines two different FD cellular network setups: one has only
uplink and downlink transmissions, the other includes D2D transmission
in addition. We consider both the discrete memoryless channel case and the
Gaussian channel case.





\begin{figure}[t]
\begin{minipage}[b]{1.0\linewidth}
\psfrag{1}[][]{\small $X^N_1$}
\psfrag{2}[][]{\small $Y^N_2$}
\psfrag{3}[][]{\small $X^N_2$}
\psfrag{4}[][]{\small $Y^N_3$}
\psfrag{5}[][t]{\small $Z^N_2$}
\psfrag{6}[][]{\small $Z^N_3$}
\psfrag{7}[][]{\small $g_{21}$}
\psfrag{8}[][b]{\small $g_{31}$}
\psfrag{9}[][]{\small $g_{22}$}
\psfrag{0}[][]{\small $g_{32}$}
\psfrag{A}[][]{\small $(m_1,m_3)$}
\psfrag{B}[][]{\small $\hat m_1$}
\psfrag{C}[][b]{\small $m_2$}
\psfrag{E}[][]{\small $(\hat m_2,\hat m_3)$}
\centering
\centerline{\includegraphics[width=7.5cm]{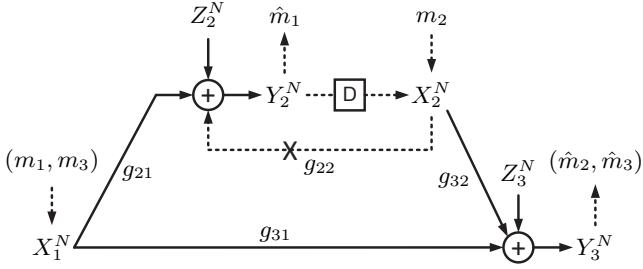}}
\caption{Gaussian full-duplex relay broadcast channel with side message $m_2$. The block ``D'' represents a {one-codeword} delay.}
\label{fig:FD_Gaussian}
\end{minipage}
\end{figure}

\subsection{Without the D2D Message}
\label{subsec:DM_nod2d}

We first consider the FD cellular network without the D2D message, as shown in Fig.~\ref{fig:model}(a).

\subsubsection{Discrete Memoryless Channel Model}
Let $X_{in}\in\mathcal X_i$ be the transmitted signal of node $i\in\{1,2\}$ and $Y_{jn}\in\mathcal Y_j$ be the received signal at node $j\in\{2,3\}$, at the $n$th channel use, over the alphabet sets $(\mathcal X_1,\mathcal X_2,\mathcal Y_2,\mathcal Y_3)$. The discrete memoryless channel model is defined by the channel transition probability mass function (pmf) $p(y_{2n},y_{3n}|x_{1n},x_{2n})$, {which captures the self-interference from $X_{1n}$ to $Y_{2n}$.} Over a total of $N$ channel uses, node 1 wishes to send $m_1\in[1:2^{NR_1}]$ to node 2, while node 2 wishes to send $m_2\in[1:2^{NR_2}]$ to node 3, where $R_1$ and $R_2$ are referred to as the uplink rate and the downlink rate, respectively. The encoding of $X_{1n}$ solely depends on $m_1$. In comparison, since the transmitter of $X_{2n}$ and the receiver of $Y_{2n}$ are co-located at node 2, the encoding of $X_{2n}$ can depend on the past received signals $Y^{n-1}_2$:
\begin{equation}
\label{encoding}
X_{1n} = \mathcal{E}_{1}(m_1,n)\;\;\mbox{and}\;\;
X_{2n} = \mathcal{E}_{2}(m_2,{Y}^{n-1}_2,n),
\end{equation}
for $n\in[1:N]$. After $N$ channel uses, node 3 decodes $m_2$ based on $Y^N_3$. Because node 2 itself is the downlink transmitter, it can make use of $m_2$ in addition to $Y^N_2$ in decoding $m_1$, i.e.,
\begin{equation}
\label{decoding}
{\hat{m}_{1} = \mathcal{D}_1({Y}^N_2,m_2)\;\;\mbox{and}\;\;
\hat{m}_{2} = \mathcal{D}_2({Y}^N_3).}
\end{equation}
An uplink-downlink rate pair $(R_1,R_2)$ is said to be achievable if there exists a set of deterministic functions $(\mathcal{E}_1,\mathcal{E}_2,\mathcal{D}_2,\mathcal{D}_3)$ such that the error probability $\mathsf{Pr}\big\{(\hat m_1,\hat m_2)\ne(m_1,m_2)\big\}$ tends to zero as $N\rightarrow\infty$.

\subsubsection{Gaussian Channel Model}
For the scalar Gaussian channel model, we have $X_{in},Y_{jn}\in\mathbb C$. We impose power constraints on $X_{in}$, i.e., $\sum^N_{n=1}|X_{in}|^2\le NP_i$, $i\in\{1,2\}$, and have
\begin{align}
Y_{2n} &= g_{21}X_{1n} + Z_{2n},
    \label{channel_model_Y2}\\
Y_{3n} &= g_{31}X_{1n} + g_{32}X_{2n} + Z_{3n}
    \label{channel_model_Y3},
\end{align}
for $n\in[1:N]$, where $g_{ji}\in\mathbb C$ is the channel gain from the transmitter node $i$ to the receiver node $j$, and $Z_{jn}\sim\mathcal{CN}(0,\sigma^2)$ for the fixed $\sigma^2>0$ is the additive white Gaussian noise at node $j$ in the $n$th channel use.

For the vector Gaussian channel model, we assume that node 1 has $L^+_1$ transmit
antennas, node 2 has $L^-_2$ receive antennas and $L^+_2$ transmit antennas,
and node 3 has $L^-_3$ receive antennas. The generalizations of $g_{ji}$, $X_{in}$, $Y_{jn}$, and $Z_{jn}$ to this vector case are $\bG_{ji}\in\mathbb
C^{L^-_{j}\times L^+_i}$, $\mathbf X_{in}\in\mathbb C^{L^+_i}$, $\mathbf Y_{jn}\in\mathbb C^{L^-_j}$, and $\mathbf Z_{jn}\in\mathbb C^{L^-_j}$, respectively. We remark that $\mathbf Z_{jn}$ is an i.i.d. vector Gaussian random variable drawn from $\mathcal{CN}(\mathbf 0,\sigma^2\bI)$. In particular, the power constraints now become $\sum^N_{n=1}\|\mathbf X_{in}\|^2\le NP_i$, $i\in\{1,2\}$. Thus, we have
\begin{align}
\mathbf Y_{2n} &= \mathbf G_{21}\mathbf X_{1n} + \mathbf Z_{2n},
    \label{MIMO_Y2}\\
\mathbf Y_{3n} &= \mathbf G_{31}\mathbf X_{1n} + \mathbf G_{32}\mathbf X_{2n} + \mathbf Z_{3n}.
    \label{MIMO_Y3}
\end{align}

{In both the scalar and the vector Gaussian cases, we assume that the channel state information (CSI), i.e., $\{g_{ji}\text{ or }\mathbf G_{ji}, \forall (i,j)\}$, is available everywhere.} Due to the fact that the BS (i.e., node 2) operates in a full-duplex mode, the self-interference at the relay has been removed implicitly, as illustrated in Fig.~\ref{fig:FD_Gaussian}. {Thus, we make an idealized assumption that the self-interference can be fully removed.}

\subsection{With the D2D Message}

Next, we consider the FD cellular network with the D2D message, as shown in Fig.~\ref{fig:model}(b).

\subsubsection{Discrete Memoryless Channel Model}
We now include a direct transmission of $m_3\in[1:2^{NR_3}]$ from node 1 to node 3 in the discrete memoryless channel model as described in Section \ref{subsec:DM_nod2d}; $R_3$ is referred to as the D2D rate. The channel setup, i.e., the alphabet sets $(\mathcal X_1,\mathcal X_2,\mathcal Y_2,\mathcal Y_3)$ and the channel transition probability $p(y_{2n},y_{3n}|x_{1n},x_{2n})$, remains the same as before. Because $m_1$ and $m_3$ are both transmitted from node 1, the encoding of $X_1$ now depends on $(m_1,m_3)$, i.e.,
\begin{equation}
\label{encoding_d2d}
X_{1n} = \mathcal{E}_{1}(m_1,m_3,n)\;\;\mbox{and}\;\;
X_{2n} = \mathcal{E}_{2}(m_2,{Y}^{n-1}_2,n).
\end{equation}
Moreover, since $m_2$ and $m_3$ are both intended for node 3, we define the decoding functions differently:
\begin{equation}
\label{decoding_d2d}
{\hat{m}_{1} = \mathcal{D}_1({Y}^N_2,m_2)\;\;\mbox{and}\;\;
(\hat{m}_{2},\hat m_3) = \mathcal{D}_2({Y}^N_3).}
\end{equation}
Similarly, a rate triple $(R_1,R_2,R_3)$ is said to be achievable if there exists a set of deterministic functions $(\mathcal{E}_1,\mathcal{E}_2,\mathcal{D}_1,\mathcal{D}_2)$ such that the probability of error, $\mathsf{Pr}\big\{(\hat m_1,\hat m_2,\hat m_3)\ne (m_1,m_2,m_3)\big\}$, tends to zero as $N\rightarrow\infty$.

\subsubsection{Gaussian Channel Model}
The scalar Gaussian channel model follows a similar fashion as in the without D2D case, except the extra message $m_3$. The channel outputs $(Y_{2n},Y_{3n})$ corresponding to the inputs $(X_{1n},X_{2n})$ are still given by (\ref{channel_model_Y2}) and (\ref{channel_model_Y3}). Again, the encoding functions in (\ref{encoding_d2d}) must satisfy the power constraints $\sum^N_{n=1}|X_{in}|^2\le NP_i$, $i\in\{1,2\}$. The vector Gaussian channel model can be extended to the D2D case similarly.

\renewcommand\arraystretch{1.6}
\begin{table*}[t]
\footnotesize
\centering
\caption{\small Proposed Coding Scheme for the Without D2D Case}
\begin{tabular}{|c||c|{c}|{c}|{c}|{c}|{c}|}
\hline
$t$ & 1 & 2 & $\cdots$ & $T-1$ & $T$\\
\hline
\hline
$X_1$ &
    $\bx^N_1(m^1_{11}|m^1_{10},1)$ & $\bx^N_1(m^2_{11}|m^2_{10},m^1_{10})$ &
    $\rightarrow$ &
    $\bx^N_1(m^{T-1}_{11}|m^{T-1}_{10},m^{T-2}_{10})$  & $\bx^N_1(1|1,m^{T-1}_{10})$\\
\hline
$Y_2$ &
    $(\hat{m}^{1}_{10},\hat m_{11}^1)$ &
    $(\hat{m}_{10}^2,\hat m_{11}^2)$ &
    $\rightarrow$ &
    $(\hat{m}_{10}^{T-1},\hat m_{11}^{T-1})$ &
    $\emptyset$\\
\hline
$X_2$ &
    $\bx^N_2(m^1_2|1)$ &
    $\bx^N_2(m^2_2|\hat{m}^1_{10})$ &
    $\rightarrow$ &
    $\bx^N_2(m^{T-1}_2|\hat{m}^{T-2}_{10})$ & $\bx^N_2(m^T_2|\hat{m}^{T-1}_{10})$\\
\hline
$Y_3$ &
    $(1,\hat m^1_2)$ &
    $(\hat{\hat m}^{1}_{10},\hat m^{2}_2)$ & $\leftarrow$ &
    $(\hat{\hat m}^{T-2}_{10},\hat m^{T-1}_2)$ &
    $(\hat{\hat m}^{T-1}_{10},\hat m^T_2)$\\
\hline
\end{tabular}
\label{tab:coding_procedure}
\end{table*}

\section{Discrete Memoryless Channel Model}
\label{sec:DM}


\subsection{Achievability for Discrete Memoryless Model without D2D}

As mentioned earlier, the cross-channel interference from node 1 to node 3 is the main bottleneck \cite{sab2014inband}. To address this issue, we use the BS (i.e., node 2) as a relay to facilitate cancelling the interfering signal at node 3. We further propose to split message $m_1$ (which causes the interference) so that node 3 can at least cancel a portion of the interference. The resulting achievable rate region is stated below.
\begin{theorem}
\label{theorem:DM:inner}
For the discrete memoryless FD cellular network without D2D, a rate pair $(R_1,R_2)$ is achievable
if it is in the convex hull of the rate regions
\begin{subequations}
\label{DM:inner}
\begin{align}
R_1 & \le I(X_1;Y_2|U,X_2),
    \label{DM:inner_R1}\\
R_2 & \le I(X_2;Y_3|U,V),
    \label{DM:inner_R2}\\
R_1+R_2 & \le I(X_1;Y_2|U,V,X_2) + I(U,V,X_2;Y_3),
    \label{DM:inner_R1R2}
\end{align}
\end{subequations}
over the joint pmf $p(u)p(v,x_1|u)p(x_2|u)$, {where the cardinalities of the auxiliary variables $U$ and $V$ can be bounded by
$|\mathcal U|\le |\mathcal X_1|\cdot |\mathcal X_2|+2$ and $|\mathcal V|\le |\mathcal X_1|\cdot |\mathcal X_2|+1$.}
\end{theorem}
\begin{IEEEproof}
Consider a total of $T$ blocks in order to carry out the block Markov coding. {We use the superscript $t\in[1:T]$ to denote the variables associated with block $t$.}

For each block $t$, split $m^t_1$ into a common-private message pair $(m^t_{10},m^t_{11})\in[1:2^{NR_{10}}]\times[1:2^{NR_{11}}]$ with $R_{10}+R_{11}=R_1$; the common message $m^t_{10}$ is {decoded} at both receivers (i.e., node 2 and node 3), while the private message $m^t_{11}$ is {decoded} at the intended receiver (i.e., node 2). For each block $t\in[1:T]$, generate the following codebooks in an i.i.d. fashion according to their respective distributions as in $p(u)p(v,x_1|u)p(x_2|u)$:
\begin{itemize}
\setlength\itemsep{0.2em}
    \item {Relay codebook} $\bu^N(m^{t-1}_{10})$;
    \item {Uplink common message codebook} $\bv^N(m^t_{10}|m^{t-1}_{10})$;
    \item {Uplink private message codebook} $\bx^N_1(m^t_{11}|m^t_{10},m^{t-1}_{10})$;
    \item {Downlink message codebook} $\bx^N_2(m^t_{2}|m^{t-1}_{10})$.
\end{itemize}
{Node 1 uses $(\bu^N,\bv^N,\bx_1^N)$ for encoding messages; node 2 uses $(\bu^N,\bx_2^N)$ for encoding, and uses $(\bu^N,\bv^N,\bx^N_1)$ for decoding; node 3 uses $(\bu^N,\bv^N,\bx^N_2)$ for decoding.}

In block $t$, knowing its past common message $m^{t-1}_{10}$, node 1 transmits
$\bx^N_1(m^t_{11}|m^t_{10},m^{t-1}_{10})$. Here, we set $m^0_{10}=0$ by convention.
Also, no message is transmitted in the last block, i.e., we set $m^T_{10}=m^T_{11}=0$.

In block $t$, after obtaining $\hat m^{t-1}_{10}$ from the previous block
$t-1$, node 2 transmits $\bx^N_2(m^t_2|\hat m^{t-1}_{10})$, and recovers $(\hat
m^{t}_{10},\hat m^{t}_{11})$ from the received signal $Y^N_2$ according to a
jointly $\epsilon$-strongly-typical set $\mathcal T^{(N)}_\epsilon$.
{Specifically, node 2 seeks a pair of $(\hat m^{t}_{10},\hat
m^{t}_{11})$ such that the corresponding codeword $\bx^N_1(\hat m^{t}_{10},\hat
m^{t}_{11})$ produces an empirical pmf $\pi(x_1,y_2)$ with
$|\pi(x_1,y_2)-p(x_1,y_2)|\le \epsilon p(x_1,y_2)$ for all
$(x_1,y_2)\in\mathcal X_1\times\mathcal Y_2$, namely \emph{strong typicality} \cite{cover2006eit}. 
}

{By the packing lemma \cite{elgamal2011nit}}, the error probability $\mathsf{Pr}\big\{(\hat m^t_{10},\hat m^t_{11})\ne (m^t_{10},m^t_{11})\big\}$ tends to zero as $N\rightarrow\infty$ provided that
\begin{align}
R_{11}&\le I(X_1;Y_2|U,V,X_2),
    \label{proof:G:R11}\\
R_{10}+R_{11}&\le I(X_1;Y_2|U,X_2).
\label{proof:G:R10R11}
\end{align}

Node 3 decodes the blocks in a \emph{backward} direction, {i.e., block $t$ prior to block $t-1$}. In block $t$, after obtaining $m^t_{10}$ from the previous block $t+1$, node 3 recovers $(\hat{\hat{m}}^{t-1}_{10},\hat{m}^t_2)$ jointly from the received signal $Y^N_2$; the error probability $\mathsf{Pr}\big\{(\hat{\hat m}^{t-1}_{10},\hat m^t_{2})\ne (m^{t-1}_{10},m^t_{2})|\hat m^{t-1}_{10}=m^{t-1}_{10}\big\}$ tends to zero as $N\rightarrow\infty$ if
\begin{align}
\label{proof:G:R2}
R_2 &\le I(X_2;Y_3|U,V),\\
\label{proof:G:R10R2}
R_{10} + R_2 &\le I(U,V,X_2;Y_3).
\end{align}

The overall error probability $P_e$, i.e., $\mathsf{Pr}\big\{({\hat m}_{1},\hat m_{2})\ne (m_{1},m_{2})\big\}$, can be upper bounded as
\begin{align}
P_e&\le\frac{1}{T}\cdot\sum^{T}_{t=1}\Big[\mathsf{Pr}\big\{(\hat m^t_{10},\hat m^t_{11})\ne (m^t_{10},m^t_{11})\big\}\,+\notag\\
&\quad\;\mathsf{Pr}\big\{(\hat{\hat m}^{t-1}_{10},\hat m^t_{2})\ne (m^{t-1}_{10},m^t_{2})\big|\hat m^{t-1}_{10}=m^{t-1}_{10}\big\}\Big],
\end{align}
so $P_e$ tends to zero as $N\rightarrow\infty$ if (\ref{proof:G:R11})--(\ref{proof:G:R10R2}) are satisfied. Furthermore, combining (\ref{proof:G:R11})--(\ref{proof:G:R10R2}) with $R_{10},R_{11}\ge0$ and $R_{1}=R_{10}+R_{11}$ by using the Fourier-Motzkin elimination, we obtain the inner bound in (\ref{DM:inner}). Note that the effective uplink rate equals to $(T-1)/T\cdot R_1$.
The achievability of (\ref{DM:inner}) is established by letting $T\rightarrow\infty$. Finally, the cardinality bounds on $U$ and $V$ are due to the property of convex set.
\end{IEEEproof}


Table \ref{tab:coding_procedure} summarizes the above coding scheme, in which the arrows show the orderings of blocks for encoding or decoding.

In Theorem \ref{theorem:DM:inner}, the auxiliary variable $U$ enables relaying at node 2 to assist node 3 in cancelling the cross-channel interference. The resulting achievable rate region can be strictly larger than that of the no relaying scheme (with $U=\emptyset$). 


\subsection{Converse for Discrete Memoryless Model without D2D}

The best previous converse is due to \cite{Tannious07}, which
proposes an outer bound for a more general model with D2D with $(R_1,R_2,R_3)$.
When specialized to the without D2D case, i.e., when $R_3=0$, their converse
amounts to the cut-set bound which consists of two individual upper bounds on
$R_1$ or $R_2$. {In this section, we propose a new upper
bound on $R_1+R_2$ that improves the cut-set bound.} This new bound is stated in the following theorem.
\begin{theorem}
\label{theorem:DM:outer}
For the discrete memoryless FD cellular network without D2D, any achievable rate pair $(R_1,R_2)$ must satisfy
\begin{subequations}
\label{DM:outer}
\begin{align}
R_1 &\le I(X_1;Y_2|X_2),
    \label{DM:outer_R1}\\
R_2 &\le I(X_2;Y_3|X_1),
    \label{DM:outer_R2}\\
R_1+R_2 &\le I(X_1;Y_2,Y_3|X_2)+I(X_2;Y_3),
    \label{DM:outer_R1R2}
\end{align}
\end{subequations}
for some joint pmf $p(x_1,x_2)$.
\end{theorem}
\begin{IEEEproof}
Let $M_i$ be the random variable denoting the message $m_i$.
Observe that (\ref{DM:outer_R1}) and (\ref{DM:outer_R2}) are simply the cut-set bounds.
{In deriving the sum-rate bound (\ref{DM:outer_R1R2}), the main idea is to introduce variable $X^N_1$ into the mutual information term $I(M_1;Y^N_2,X^N_2)$, as in the following:}
\allowdisplaybreaks
\begin{align}
&N(R_1+R_2-\epsilon_N)\notag\\
&\le I(M_1;Y^N_2,X^N_2) + I(M_2; Y^N_3)\notag\\
&\overset{(a)}\le I(M_1;Y^N_2,X^N_2,Y^N_3|M_2) + I(M_2;Y^N_3)\notag\\
&\overset{(b)}= I(M_1;Y^N_2,Y^N_3|M_2) + I(M_2;Y^N_3)\notag\\
&\le \sum^N_{n=1}\Big[I(M_1;Y_{2n},Y_{3n}|M_2,Y^{n-1}_2,Y^{n-1}_3)\notag\\ &\;\quad+I(M_2,Y^{n-1}_2,Y^{n-1}_3;Y_{3n})\Big]\notag\\
&\overset{(c)}= \sum^N_{n=1}\Big[I(M_1;Y_{2n},Y_{3n}|M_2,Y^{n-1}_2,Y^{n-1}_3,X_{2n})\notag\\ &\;\quad+I(M_2,Y^{n-1}_2,Y^{n-1}_3,X_{2n};Y_{3n})\Big]\notag\\
&\le\sum^N_{n=1}\Big[I(M_1,M_2,Y^{n-1}_2,Y^{n-1}_3;Y_{2n},Y_{3n}|X_{2n})\notag\\
&\;\quad+I(X_{2n};Y_{3n})\Big]\notag\\
&\overset{(d)}=\sum^N_{n=1}\Big[I(X_{1n};Y_{2n},Y_{3n}|X_{2n})+I(X_{2n};Y_{3n})\Big]\notag\\
&\le NI(X_1;Y_2,Y_3|X_2) + NI(X_2;Y_3),
    \label{DM_proof:outer_R1R2}
\end{align}
where $\epsilon_N$ tends to zero as $N\rightarrow\infty$ by Fano's inequality,
$(a)$ follows as $M_1\indep M_2$, $(b)$ and $(c)$ both follow as $X_{2n}$ is a deterministic function of $(M_2,Y^{n-1}_2)$, $(d)$ follows as $(M_1,M_2,Y^{n-1}_2,Y^{n-1}_3)\rightarrow X_{1n}\rightarrow (Y_{2n},Y_{3n})$ form a Markov chain conditioned on $X_{2n}$. The converse is then verified.
\end{IEEEproof}

{Note that we restrict the relay operation to be deterministic in the channel model and in deriving the above outer bound. We remark that (\ref{DM:outer_R1R2}) is not contained in the cut-set bound and yet is critical to
characterizing the capacity region to within a constant gap for the scalar Gaussian
channel in Section \ref{subsec:gap}.}

\subsection{Achievability for Discrete Memoryless Model with D2D}


\renewcommand\arraystretch{1.6}
\begin{table*}[t]
\footnotesize
\centering
\caption{Proposed Coding Scheme for the D2D Case With D2D Rate Splitting}
\begin{tabular}{|c|c|{c}|{c}|{c}|{c}|}
\hline
$t$ & 1 & 2 & $\cdots$ & $T-1$ & $T$\\
\hline
\hline
$X_1$ &
    $\bx^N_1(m^1_1,m^1_{30},m^1_{33}|1,1)$ & $\bx^N_1(m^2_1,m^2_{30},m^2_{33}|m^1_1,m^1_{30})$ &
    $\rightarrow$ &
    $\bx^N_1(m^{T-1}_1,m^{T-1}_{30},m^{T-1}_{33}|m^{T-2}_1,m^{T-2}_{30})$  & $\bx^N_1(1,1,1|m^{T-1}_1,m^{T-1}_{30})$\\
    \hline
$Y_2$ &
    $(\hat{m}_{1}^1,\hat{\hat m}_{30}^1,\hat{\hat m}^1_{33})$ &
    $(\hat{m}_{1}^2,\hat{\hat m}_{30}^2,\hat{\hat m}^2_{33})$ &
    $\rightarrow$ &
    $(\hat{m}_{1}^{T-1},\hat{\hat m}_{30}^{T-1},\hat{\hat m}^{T-1}_{33})$ &
    $\emptyset$\\
    \hline
$X_2$ &
    $\bx^N_2(m^1_2|1,1)$ & $\bx^N_2(m^2_2|\hat{m}^1_1,\hat{\hat m}^1_{30})$ &
    $\rightarrow$ &
    $\bx^N_2(m^{T-1}_2|\hat{m}^{T-2}_1,\hat{\hat m}^{T-2}_{30})$ & $\bx^N_2(m^T_2|\hat{m}^{T-1}_1,\hat{\hat m}^{T-1}_{30})$\\
    \hline
$Y_3$ &
    $(1,\hat m^1_2,1,\hat m^1_{33})$ &
    $(\hat{\hat m}^{1}_1,\hat m^{2}_2,\hat m_{30}^{1},\hat m^{2}_{33})$ &
    $\leftarrow$ &
    $(\hat{\hat m}^{T-2}_1,\hat m^{T-1}_2,\hat m_{30}^{T-2},\hat m^{T-1}_{33})$ &
    $(\hat{\hat m}^{T-1}_1,\hat m^T_2,\hat m^{T-1}_{30},1)$\\
\hline
\end{tabular}
\label{tab:coding_procedure_d2d:m3}
\end{table*}

Recall that the FD cellular network with D2D is a generalization of the FD relay broadcast channel \cite{LiangTIT2007a,LiangTIT2007b}, so
we use the previous studies in \cite{LiangTIT2007a,LiangTIT2007b} as a starting point.
The works \cite{LiangTIT2007a,LiangTIT2007b} propose to modify the classic
Marton's coding \cite{MartonTIT1979} for the broadcast channel to the case where one receiver further
helps the other receiver via a relay link. The channel model considered in this
paper is a further generalization in which the extra side message $m_2$ is
carried in this relay link. The coding strategy proposed below incorporates
$m_2$ in Marton's broadcast coding.

The coding strategy of \cite{LiangTIT2007a,LiangTIT2007b} splits each message
(i.e., $m_1$ and $m_3$) into the private and common parts which are dealt with
differently.  The common part is decoded by both node 2 and node 3; node 2
further acts as a relay to assist node 3 in decoding the common message. In
contrast, the private parts are decoded only by the intended node through the
broadcast channel without using node 2 as relay, so Marton's coding can be
applied.  This paper makes two modifications to this strategy in order to
enable an extra transmission of $m_2$.  First, we let the encoding of $X_2$ be based on
both $m_1$ and $m_2$. Second, we let node 3 decode the original common and
private message jointly with the new message $m_2$. The resulting achievable
rate region is stated as follows.
\begin{theorem}
\label{theorem:DM_d2d:inner}
For the discrete memoryless FD cellular network with D2D, a rate triple $(R_1,R_2,R_3)$ is achievable
if it is in the convex hull of the rate regions
\allowdisplaybreaks
\begin{subequations}
\label{DM_d2d:inner}
\begin{align}
R_1 &\le \mu_3,\\
R_2 &\le \min\{\mu_5,\mu_2+\mu_6-\mu_1\},\\
R_1+R_3 &\le \mu_3+\mu_4-\mu_1,\\
R_2+R_3 &\le \mu_7,\\
R_1+R_2+R_3 &\le \min\{\mu_2+\mu_7-\mu_1,\mu_3+\mu_6-\mu_1\},
\end{align}
\end{subequations}
over the joint pmf $p(u)p(v,w_1,w_3,x_1|u)p(x_2|u)$ under the constraint that $\mu_1 \le \mu_2+\mu_4$, where
\begin{subequations}
\label{Marton:mutual_info}
\begin{align}
\mu_1 &= I(W_1;W_3|U,V),\\
\mu_2 &= I(W_1;Y_2|U,V,X_2),\\
\mu_3 &= I(V,W_1;Y_2|U,X_2),\\
\mu_4 &= I(W_3;Y_3|U,V,X_2),\\
\mu_5 &= I(X_2;Y_3|U,V,W_3),\\
\mu_6 &= I(W_3,X_2;Y_3|U,V),\\
\mu_7 &= I(U,V,W_3,X_2;Y_3).
\end{align}
\end{subequations}
\end{theorem}
\begin{IEEEproof}
Again, we consider a total of $T$ blocks and {use $t\in[1:T]$ to index the block.} For each block $t$, split $m^t_i$ into the common-private message pair
$(m^t_{i0},m^t_{ii})\in[1:2^{nR_{i0}}]\times[1:2^{NR_{ii}}]$ for $i\in\{1,3\}$. For each block
$t$, in an i.i.d. manner according to their respective distributions,
generate a common codebook
\begin{itemize}
\setlength\itemsep{0.2em}
    \item {Relay codebook} $\bu^N(m^{t-1}_{10},m^{t-1}_{30})$;
    \item {Common codebook} $\bv^N(m^t_{10},m^t_{30}|m^{t-1}_{10},m^{t-1}_{30})$;
    \item Separate binning codebooks $\mathbf w^N_1(\ell^t_{11})$ and $\mathbf w^N_3(\ell^t_{33})$;
    \item {Joint binning codebook} $\mathbf x^N_1(\ell^t_{11},\ell^t_{33}|m_{10}^{t-1},m_{30}^{t-1})$;
    \item {Downlink codebook} $\mathbf x^N_2(m^t_2|m^{t-1}_{10},m^{t-1}_{30})$,
\end{itemize}
where the codebook pair $\big(\mathbf w^N_1(\ell_{11}),\mathbf w^N_3(\ell_{33})\big)$ is generated for each $(\ell_{11},\ell_{33})\in[1:2^{NR'_{11}}]\times [1:2^{NR'_{11}}]$, with $R'_{ii}\ge R_{ii}$, $i\in\{1,3\}$, and with each $\ell^t_{ii}$ uniformly mapped to the bin of $m^t_{ii}$, namely $\mathcal B_i(m^t_{ii})$. {Node 1 uses $(\bu^N,\bv^N,\bw^N_1,\bw^N_3,\bx^N_1)$ for encoding; node 2 uses $(\bu^N,\bx^N_2)$ for encoding, and uses $(\bu^N,\bv^N,\bw^N_1,\bx^N_1)$ for decoding; node 3 uses $(\bu^N,\bw^N_3,\bx^N_2)$ for decoding.}

In block $t$, node 1 finds a pair of $(\ell^t_{11},\ell^t_{33})\in\mathcal B_1(m^t_{11})\times\mathcal B_3(m^t_{33})$ such that $(\mathbf w^N_1(\ell^t_{11}),\mathbf w^N_3(\ell^t_{33}))$ is in a strongly typical set $\mathcal T^{(N)}_{\epsilon'}$, then transmits $\mathbf x^N_1(\ell^t_{11},\ell^t_{33}|m_{10}^{t-1},m_{30}^{t-1})$. {The above typicality criterion $\mathcal T^{(N)}_{\epsilon'}$ needs to be stricter than the typicality criterion $\mathcal T^{(N)}_\epsilon$ used for decoding in the sense that $0 < \epsilon' < \epsilon$.} This encoding is guaranteed to be successful provided that
\begin{equation}
\label{inner:ineq1}
R'_{11}+R'_{33} - R_{11}-R_{33} \ge I(W_1;W_3|U,V).
\end{equation}

In block $t$, after obtaining $(\hat m^{t-1}_{10},\hat{\hat{m}}^{t-1}_{30})$ from the previous block $t-1$, node 2 transmits $\mathbf x^N_2(m^t_2|\hat m^{t-1}_{10},\hat{\hat{m}}^{t-1}_{30})$, and recovers $(\hat m^{t}_{10}, \hat{\hat{m}}^{t}_{30})$ jointly from the received signal $Y^N_2$; this decoding is successful if
\begin{align}
R'_{11} &\le I(W_1;Y_2|U,V,X_2),
    \label{inner:ineq2}\\
R_{10}+R_{30}+R'_{11} &\le I(V,W_1;Y_2|U,X_2).
    \label{inner:ineq3}
\end{align}

Node 3 decodes the blocks in a backward direction (unlike the sliding window decoding scheme of \cite{LiangTIT2007b}), i.e., block $t$ prior to block $t-1$. In block $t$, after obtaining $(\hat{\hat{m}}^t_{10},\hat m^t_{30})$ from the previous block $t+1$, node 1 recovers $(\hat{\hat{m}}^t_{10},\hat m^t_{30},\hat m^t_{33},\hat m^t_2)$ jointly; the following conditions guarantee successful decoding:
\begin{align}
R'_{33} &\le I(W_3;Y_3|U,V,X_2),
    \label{inner:ineq4}\\
R_2 &\le I(X_2;Y_3|U,V,W_3),
    \label{inner:ineq5}\\
R'_{33}+R_2 &\le I(W_3,X_2;Y_3|U,V),
    \label{inner:ineq6}\\
R_{10}+R_{30}+R'_{33}+R_2 &\le I(U,V,W_3,X_2;Y_3).
    \label{inner:ineq7}
\end{align}
Combining (\ref{inner:ineq1})--(\ref{inner:ineq7}) with $R_{11}\le R'_{11}$, $R_{33}\le R'_{33}$, $R_1=R_{10}+R_{11}$, $R_3=R_{30}+R_{33}$, and a nonnegative constraint on all the rate variables,
and letting $T\rightarrow\infty$,
we establish the proposed inner bound, including the constraint
$\mu_1\le\mu_2+\mu_4$, via the Fourier-Motzkin elimination.
\end{IEEEproof}

\begin{remark}
The inner bound in Theorem \ref{theorem:DM_d2d:inner} reduces to that of \cite{LiangTIT2007a,LiangTIT2007b} for the FD relay broadcast channel when $U=X_2$, reduces to a decode-and-forward inner bound (\ref{DM:inner}) of \cite{sab2014inband} for the D2D case when $W_1=W_3=\emptyset$, and reduces to the inner bound in Theorem \ref{theorem:DM:inner} for the without D2D case when $W_3=\emptyset$.
\end{remark}

\begin{remark}
The earlier works \cite{LiangTIT2007a,LiangTIT2007b} also use the decode-and-forward relaying but
with sliding window decoding. This work uses backward decoding.
\end{remark}

In Theorem \ref{theorem:DM_d2d:inner}, the term $\mu_1$ is due to Marton's
coding \cite{MartonTIT1979}, reflecting the extent to which the encodings of the private messages
$m_{11}$ and $m_{33}$ are coordinated through broadcasting. 
The following proposition further shows that the constraint $\mu_1\le\mu_2+\mu_4$ must be satisfied
automatically if $p(u)p(v,w_1,w_3,x_1|u)p(x_2|u)$ is optimally chosen for
maximizing the rate region (\ref{DM_d2d:inner}).

\begin{proposition}
\label{prop:remove_cond}
The achievable rate region of Theorem \ref{theorem:DM_d2d:inner} remains the same if the constraint $\mu_1\le\mu_2+\mu_4$ is removed.
\end{proposition}
\begin{IEEEproof}
Let $\mathcal R_1\subseteq\mathbb R^3_+$ be the achievable rate region defined by the set of inequalities in Theorem
\ref{theorem:DM_d2d:inner}, and let $\mathcal R_2$ be the version without the
constraint $\mu_1\le\mu_2+\mu_4$. Clearly, $\mathcal R_1\subseteq \mathcal R_2$, so it suffices to show $\mathcal R_2\subseteq\mathcal R_1$ in order to prove $\mathcal R_1=\mathcal R_2$. Consider some $p(u)p(v,w_1,w_3,x_1|u)p(x_2|u)$ such that $\mu_1>\mu_2+\mu_4$. Under this pmf, it can be shown that $\mathcal R_2\subseteq\mathcal R'_2$ where $\mathcal R'_2$ is another rate region defined by
\begin{subequations}
\begin{align}
R_2 &\le \min\{\mu_5,I(X_2;Y_3|U,V)\},\\
R_1+R_3 &\le I(V;Y_2|U,X_2),\\
R_1+R_2+R_3 &\le I(U,V,X_2;Y_3).
\end{align}
\end{subequations}
In the meanwhile, $\mathcal R'_2$ can be attained by setting $W_1=\emptyset$
in Theorem \ref{theorem:DM_d2d:inner}. Thus, $\mathcal R_2\subseteq\mathcal R_1$.
\end{IEEEproof}

\begin{table*}[t]
\footnotesize
\centering
\caption{Proposed Coding Scheme for the D2D Case With Uplink Rate Splitting}
\begin{tabular}{|c|c|{c}|{c}|{c}|{c}|}
\hline
$t$ & 1 & 2 & $\cdots$ & $T-1$ & $T$\\
\hline
\hline
$X_1$ &
    $\bx^N_1(m^1_{10},m^1_{11},m^1_{3}|1,1)$ & $\bx^N_1(m^2_{10},m^2_{11},m^2_{3}|m^1_{10},m^1_{3})$ &
    $\rightarrow$ &
    $\bx^N_1(m^{T-1}_{10},m^{T-1}_{11},m^{T-1}_{3}|m^{T-2}_{10},m^{T-2}_{3})$  & $\bx^N_1(1,1,1|m^{T-1}_{10},m^{T-1}_{3})$\\
    \hline
$Y_2$ &
    $(\hat{m}_{1}^{10},\hat m_{11}^1,\hat{\hat m}^1_{3})$ &
    $(\hat{m}_{10}^2,\hat m_{11}^2,\hat{\hat m}^2_{3})$ &
    $\rightarrow$ &
    $(\hat{m}_{10}^{T-1},\hat m_{11}^{T-1},\hat{\hat m}^{T-1}_{3})$ &
    $\emptyset$\\
    \hline
$X_2$ &
    $\bx^N_2(m^1_2|1,1)$ &
    $\bx^N_2(m^2_2|\hat{m}^1_{10},\hat{\hat m}^1_{3})$ &
    $\rightarrow$ &
    $\bx^N_2(m^{T-1}_2|\hat{m}^{T-2}_{10},\hat{\hat m}^{T-2}_{3})$ & $\bx^N_2(m^T_2|\hat{m}^{T-1}_{10},\hat{\hat m}^{T-1}_{3})$\\
    \hline
$Y_3$ &
    $(1,\hat m^1_2,1)$ &
    $(\hat{\hat m}^{1}_{10},\hat m^{2}_2,\hat m^{1}_{3})$ & $\leftarrow$ &
    $(\hat{\hat m}^{T-2}_{10},\hat m^{T-1}_2,\hat m^{T-2}_{3})$ &
    $(\hat{\hat m}^{T-1}_{10},\hat m^T_2,\hat m^{T-1}_{3})$\\
\hline
\end{tabular}
\label{tab:coding_procedure_d2d:m1}
\end{table*}

The inner bound in Theorem \ref{theorem:DM_d2d:inner} involves rate splitting for both $m_1$ and $m_3$. The following two corollaries present the special cases in which only one of $(m_1,m_3)$ has rate splitting and Marton's coding is replaced with the superposition coding.
\begin{corollary}[D2D Rate Splitting]
\label{theorem:DM_d2d:inner1}
For the discrete memoryless FD cellular network with D2D, a rate triple $(R_1,R_2,R_3)$ is achievable
if it is in the convex hull of
\begin{subequations}
\label{DM_d2d:inner1}
\begin{align}
R_1 &\le I(V;Y_2|U,X_2),\\
R_2 &\le I(X_2;Y_3|U,X_1),\\
R_1+R_3 &\le I(V;Y_2|U,X_2)\notag\\
&\quad\, +I(X_1;Y_3|U,V,X_2),\\
R_1+R_2+R_3 &\le \min\{I(V;Y_2|U,X_2)+I(X_1,X_2;Y_3|U,V),\notag\\
&\qquad\qquad I(X_1,X_2;Y_3)\},
\end{align}
\end{subequations}
over the joint pmf $p(u)p(v,x_1|u)p(x_2|u)$, {where the cardinalities of auxiliary variables can be bounded by $|\mathcal U|\le |\mathcal X_1|\cdot|\mathcal X_2|+3$ and $|\mathcal V|\le|\mathcal X_1|+2$.}
\end{corollary}
\begin{IEEEproof}
This inner bound is obtained by setting $W_1=\emptyset$ and $W_3=X_1$ in (\ref{DM_d2d:inner}) of Theorem \ref{theorem:DM_d2d:inner}. The corresponding encoding and decoding procedure is illustrated in {Table \ref{tab:coding_procedure_d2d:m3}}. 
\end{IEEEproof}

\begin{corollary}[Uplink Rate Splitting]
\label{theorem:DM_d2d:inner2}
For the discrete memoryless FD cellular network with D2D, a rate triple $(R_1,R_2,R_3)$ is achievable if
it is in the convex hull of
\begin{subequations}
\label{DM_d2d:inner2}
\begin{align}
R_2 &\le I(X_2;Y_3|U,V),\\
R_1+R_3 &\le I(X_1;Y_2|U,X_2),\\
R_2+R_3 &\le I(U,V,X_2;Y_3),\\
R_1+R_2+R_3 &\le I(X_1;Y_2|U,V,X_2)\notag\\
&\quad\, +I(U,V,X_2;Y_3),
\end{align}
over the joint pmf $p(u)p(v,x_1|u)p(x_2|u)$, {where the cardinalities of the auxiliary variables can be bounded by $|\mathcal U|\le |\mathcal X_1|\cdot|\mathcal X_2|+3$ and $|\mathcal V|\le|\mathcal X_1|+2$.}
\end{subequations}
\end{corollary}
\begin{IEEEproof}
This inner bound is obtained by setting $W_3=\emptyset$ and $W_1=X_1$ in (\ref{DM_d2d:inner}) of Theorem \ref{theorem:DM_d2d:inner}. The corresponding encoding and decoding procedure is illustrated in Table \ref{tab:coding_procedure_d2d:m1}. 
\end{IEEEproof}

\begin{remark}
\label{remark:uplink_rate_splitting_reduce_to_without_D2D}
The inner bound (\ref{DM_d2d:inner2}) reduces to the previous inner bound (\ref{DM:inner}) for the without D2D case when $R_3=0$.
\end{remark}

It turns out that using one of the two special cases according to the channel condition can already achieve the capacity to within a constant gap for the scalar Gaussian case, as shown in Section \ref{subsec:DM_d2d:outer}

\subsection{Converse for Discrete Memoryless Model with D2D}
\label{subsec:DM_d2d:outer}

The existing works \cite{LiangTIT2007a,LiangTIT2007b} on the FD relay broadcast channel use
auxiliary ``{genie}'' variables to improve the cut-set bound. Similarly, with the aid of genie, \cite{Tannious07} enhances the cut-set bound for the case with relay-to-destination side message.
As compared to \cite{Tannious07}, we provide two improvements. First, we further tighten the
genie-aided bound by using more suitable auxiliary
variables. Second, we propose a new upper bound on $R_1+R_2+R_3$ that improves
the cut-set bound. Our converse is specified in the following.

\begin{theorem}
\label{theorem:DM_d2d:outer}
For the discrete memoryless FD cellular network with D2D, any achievable rate triple $(R_1,R_2,R_3)$ must satisfy
\begingroup
\allowdisplaybreaks
\begin{subequations}
\label{DM_d2d:outer}
\begin{align}
R_1 &\le \min\{I(U;Y_2|X_2),\notag\\
    &\qquad\qquad I(X_1;Y_2,Y_3|V,X_2)\},
    \label{DM_d2d:outer:aa}\\
R_2 &\le I(X_2;Y_3|X_1),
    \label{DM_d2d:outer:bb}\\
R_3 &\le \min\{I(X_1;Y_2,Y_3|U,X_2),\notag\\
    &\qquad\qquad I(V;Y_2,Y_3|X_2)\},
    \label{DM_d2d:outer:cc}\\
R_1+R_3 &\le I(X_1;Y_2,Y_3|X_2),
    \label{DM_d2d:outer:dd}\\
R_2+R_3 &\le I(X_1,X_2;Y_3),
    \label{DM_d2d:outer:ee}\\
R_1+R_2+R_3 &\le I(X_1;Y_2,Y_3|X_2) + I(X_2;Y_3),
    \label{DM_d2d:outer:ff}
\end{align}
\end{subequations}
for some joint pmf $p(u,v,x_1,x_2)$, {where the cardinalities of the auxiliary variables can be bounded by $|\mathcal U|\le |\mathcal X_1|\cdot|\mathcal X_1|+1$ and $|\mathcal V|\le |\mathcal X_1|\cdot|\mathcal X_1|+1$.}
\end{theorem}
\endgroup
\begin{IEEEproof}
Observe that those bounds in (\ref{DM_d2d:outer:aa})--(\ref{DM_d2d:outer:ee}) without $U$ or $V$ are directly from the cut-set bound. The existing work \cite{Tannious07} assumes a genie that provides $U'_n=(Y^{n-1}_2,Y^{n-1}_3)$ and $V'_n=M_3$ to node 1 and node 2.
In contrast, we propose a different genie that provides
$U_n = (M_1,M_2,Y^{n-1}_2,Y^{n-1}_3)$ to \mbox{node} 1 and node 3,
and provides $V_n = (M_2,M_3,Y^{n-1}_2,Y^{n-1}_3)$
to node 1 and node 2. This new use of genie yields a tighter outer bound.

The upper bound (\ref{DM_d2d:outer:ff}) on the sum rate is obtained as follows. Considering node 2 and node 3 as two receivers, we follow Sato's
approach in \cite{satoIT78} and assume that they could fully coordinate in their
decoding with the aid of genie. Considering node 2 as the transmitter of $m_2$, we introduce a genie
that provides feedback $\bY^{n-1}_3$ to it to improve encoding.
The converse is then established by letting $N\rightarrow\infty$. The complete proof is shown in Appendix \ref{proof:DM_d2d:outer}.
\end{IEEEproof}

\begin{remark}
In contrast to the previous converse in \cite{Tannious07}, which is not computable, the converse of Theorem
\ref{theorem:DM_d2d:outer} can be evaluated. We do so for the Gaussian case in the next section. 
\end{remark}

\section{Scalar Gaussian Channel Model}
\label{sec:scalar_G} 

{The mutual information bounds for the discrete memoryless channel model can be carried over to the Gaussian case. We now evaluate the achievability and converse for the scalar Gaussian channel model under power constraints}. 

\subsection{Achievability for Scalar Gaussian Model}

We first compute the mutual information inner bound (\ref{DM:inner}).
\begin{proposition}
\label{prop:G:inner}
For the scalar Gaussian FD cellular network without D2D, a rate pair $(R_1,R_2)$ is achievable if it is in the convex hull of
\begin{subequations}
\label{G:inner}
\begin{align}
R_1 & \le \mathsf C\bigg(\frac{(b+c)|g_{21}|^2P_1}{\sigma^2} \bigg),
    \label{G:inner_R1}\\
R_2 & \le \mathsf C\bigg(\frac{e|g_{32}|^2P_2}{\sigma^2+c |g_{31}|^2P_1}\bigg),
    \label{G:inner_R2}\\
R_1+R_2 & \le \mathsf C\bigg(\frac{(a+b)|g_{31}|^2P_1+|g_{32}|^2P_2+J\sqrt{ad}}{\sigma^2+c |g_{31}|^2P_1}\bigg) \notag\\
&\quad\,+\mathsf C\bigg(\frac{c |g_{21}|^2P_1}{\sigma^2}\bigg),
    \label{G:inner_R1R2}
\end{align}
\end{subequations}
over the nonnegative parameters $(a,b,c,d,e)$ with $a+b+c\le1$
and $d+e\le1$, where
\begin{equation}
\label{def_J}
J=2|g_{31}g_{32}|\sqrt{P_1P_2}.
\end{equation}
\end{proposition}
\begin{IEEEproof}
{Generate the codebooks $\mathbf{u}^{N}(m^{t-1}_{10})$, $\tilde{\mathbf{v}}^{N}(m^t_{1})$, $\mathbf{w}^{N}_{1}(m^t_{11})$, and $\mathbf{w}^{N}_2(m^t_{2})$, all according to $\mathcal{CN}(0,1)$ in an i.i.d. fashion. In block $t\in[1:T]$, node 1 transmits
\begin{equation}
\bx^N_1(t) = \bv^N(t)+\sqrt{cP_1}\bw^N_1(m^t_{11}),
\end{equation}
where
\begin{equation}
\bv^N(t) = \sqrt{aP_1}\bu^N(m^{t-1}_{10})+\sqrt{bP_1}\tilde\bv^N(m^t_{10}).
\end{equation}
In block $t$, upon learning $\hat m^{t-1}_{10}$ from the previous block $t-1$, node 2 transmits
\begin{equation}
\bx^N_2(t) = \sqrt{dP_2}\bu^N(\hat{m}^{t-1}_{10})+
\sqrt{eP_2}\bw^N_2(m^t_2).
\end{equation}
Plugging the above setting into (\ref{DM:inner}) gives (\ref{G:inner}). Furthermore, a convex hull is obtained by time sharing across different choices of $(a,b,c,d,e)$.}
\end{IEEEproof}

{
\begin{remark}
We remark that an alternative way to achieve the inner bound (\ref{G:inner}) is
by using the binning strategy, i.e., by partitioning $m^t_{10}$ into bins. The
bin index is transmitted by node 1, then relayed by node 2, and used by node 3
in the decoding of $m^t_{10}$. This binning scheme can be realized in practice
by means of hybrid automatic repeat request (HARQ) \cite{RazaghiTIT2007} in
which the bin index is basically the parity bits of $m^t_{10}$.
\end{remark}}

We show that the above achievability coincides with the capacity under a \emph{very strong interference} regime.

\begin{theorem}
\label{prop:G_very_strong:capacity}
For the scalar Gaussian FD cellular network without D2D, in the \emph{very strong interference} regime, i.e., when $|g_{31}|^2\ge
|g_{21}|^2(1+|g_{32}|^2)$, the capacity region of the rate pair $(R_1,R_2)$ is
\begin{subequations}
\label{G_very_strong:inner}
\begin{align}
R_1 & \le \mathsf C\bigg(\frac{|g_{21}|^2P_1}{\sigma^2} \bigg),
    \label{}\\
R_2 & \le \mathsf C\bigg(\frac{|g_{32}|^2P_2}{\sigma^2}\bigg).
    \label{}
\end{align}
\end{subequations}
\end{theorem}
\begin{IEEEproof}
The achievability is verified directly by setting $a=c=d=0$ and $b=e=1$ in (\ref{G:inner}); note that (\ref{G:inner_R1R2}) becomes redundant if $|g_{31}|^2\ge
|g_{21}|^2(1+|g_{32}|^2)$. The converse directly follows from the cut-set bound.
\end{IEEEproof}


We now consider the D2D case. Instead of the full set of mutual information bounds in Theorem \ref{theorem:DM_d2d:inner}, we evaluate the two simpler inner bounds in (\ref{DM_d2d:inner1}) and (\ref{DM_d2d:inner2}), as in the following.


\begin{proposition}[D2D Rate Splitting]
\label{prop:G_d2d:inner1}
For the scalar Gaussian FD cellular network with D2D, a rate triple $(R_1,R_2,R_3)$ is achievable if it satisfies
\allowdisplaybreaks
\begin{subequations}
\label{G_d2d:inner1}
\begin{align}
R_1 &\le \mC\bigg(\frac{b|g_{21}|^2P_1}{\sigma^2+c|g_{21}|^2P_1}\bigg),\\
R_2 &\le \mC\big({e|g_{32}|^2P_2}/{\sigma^2}\big),\\
R_1+R_3 &\le \mC\bigg(\frac{b|g_{21}|^2P_1}{\sigma^2+c|g_{21}|^2P_1}\bigg)+
    \mC\bigg(\frac{c|g_{31}|^2P_1}{\sigma^2}\bigg),\\
R_1+R_2+R_3 &\le \min\bigg\{\mC\bigg(\frac{|g_{31}|^2P_1+|g_{32}|^2P_2+J\sqrt{ad}}{\sigma^2}\bigg),\notag\\
&\qquad\qquad\mC\bigg(\frac{c|g_{31}|^2P_1+e|g_{32}|^2P_2}{\sigma^2}\bigg)\notag\\
&\qquad\qquad+\mC\bigg(\frac{b|g_{21}|^2P_1}{\sigma^2+c|g_{21}|^2P_1}\bigg)\bigg\},
\end{align}
\end{subequations}
for some nonnegative parameters $(a,b,c,d,e)$ {with $a+b+c\le1$ and $d+e\le1$,} where $J$ is defined in (\ref{def_J}).
\end{proposition}
\begin{IEEEproof}
Generate the codebooks $\mathbf{w}^{N}_{2}(m^t_2)$, $\mathbf{w}^{N}_{3}(m^t_{33})$, $\tilde{\mathbf{v}}^{N}(m^t_{1},m^t_{30})$, and $\mathbf{u}^{N}(m^{t-1}_{1},m^t_{30})$, all according to $\mathcal{CN}(0,1)$ in an i.i.d. fashion. In block $t\in[1:T]$, node 1 transmits
\begin{equation}
\bx^N_1(t) = \bv^N(t)+\sqrt{cP_1}\bw^N_3(m^t_{33}),
\end{equation}
where
\begin{equation}
\bv^N(t) = \sqrt{aP_1}\bu^N(m^{t-1}_{1},m^{t-1}_{30})+\sqrt{bP_1}\tilde\bv^N(m^t_{1},m^t_{30}).
\end{equation}
In block $t$, with $(\hat m^{t-1}_1,\hat{\hat{m}}^{t-1}_{30})$ obtained from the previous block $t-1$, node 2 transmits
\begin{equation}
\bx^N_2(t) = \sqrt{dP_2}\bu^N(\hat{m}^{t-1}_{1},\hat{\hat{m}}^{t-1}_{30})+
\sqrt{eP_2}\bw^N_2(m^t_2).
\end{equation}
Substituting the above setting in (\ref{DM_d2d:inner1}) gives
(\ref{G_d2d:inner1}).
\end{IEEEproof}

The mutual information inner bound (\ref{DM_d2d:inner2}) can be computed similarly, as stated below without proof.

\begin{proposition}[Uplink Rate Splitting]
\label{prop:G_d2d:inner2}
For the scalar Gaussian FD cellular network with D2D, a rate triple $(R_1,R_2,R_3)$ is achievable if it satisfies
\allowdisplaybreaks
\begin{subequations}
\label{G_d2d:inner2}
\begin{align}
R_2 &\le \mC\bigg(\frac{e|g_{32}|^2P_2}{\sigma^2+c|g_{31}|^2P_1}\bigg),\\
R_1+R_3 &\le \mC\bigg(\frac{(b+c)|g_{21}|^2P_1}{\sigma^2}\bigg),\\
R_2+R_3 &\le \mC\bigg(\frac{(a+b)|g_{31}|^2P_1+|g_{32}|^2P_2+J\sqrt{ad}}{\sigma^2+c|g_{31}|^2P_1}\bigg),\\
R_1+R_2+R_3 &\le \mC\bigg(\frac{(a+b)|g_{31}|^2P_1+|g_{32}|^2P_2+J\sqrt{ad}}{\sigma^2+c|g_{31}|^2P_1}\bigg)\notag\\
&\quad\,+\mC\bigg(\frac{c|g_{21}|^2P_1}{\sigma^2}\bigg),
\end{align}
\end{subequations}
for some nonnegative parameters $(a,b,c,d,e)$ {with $a+b+c\le1$ and $d+e\le1$,} where $J$ is defined in (\ref{def_J}).
\end{proposition}

Although the above two inner bounds are two special cases of Theorem \ref{theorem:DM_d2d:inner}, it turns out that they suffice to attain the capacity of the scalar Gaussian channel case to within one bit, as shown in Section \ref{subsec:gap}.

\setcounter{equation}{45}
\begin{figure*}
\begin{subequations}
\label{G_d2d:outer}
\begin{align}
R_1 &\le \min\bigg\{\mC\bigg(\frac{(1-\rho^2)|g_{21}|^2P_1}{\sigma^2+\alpha(1-\rho^2)|g_{21}|^2P_1}\bigg),\mC\bigg(\frac{\beta(1-\rho^2)(|g_{21}|^2+|g_{31}|^2)P_1}{\sigma^2}\bigg)\bigg\},
    \label{G_d2d:outer:R1}\\
R_2 &\le \mC\bigg(\frac{(1-\rho^2)|g_{32}|^2P_2}{\sigma^2}\bigg),
    \label{G_d2d:outer:R2}\\
R_3 &\le \min\bigg\{\mC\bigg(\frac{\alpha(1-\rho^2)(|g_{21}|^2+|g_{31}|^2)P_1}{\sigma^2}\bigg),\mC\bigg(\frac{(1-\rho^2)(|g_{21}|^2+|g_{31}|^2)P_1}{\sigma^2+\beta(1-\rho^2)(|g_{21}|^2+|g_{31}|^2)P_1}\bigg)\bigg\},
    \label{G_d2d:outer:R3}\\
R_1+R_3 &\le \mC\bigg(\frac{(1-\rho^2)(|g_{21}|^2+|g_{31}|^2)P_1}{\sigma^2}\bigg),
    \label{G_d2d:outer:R1R3}\\
R_2+R_3 &\le \mC\bigg(\frac{|g_{31}|^2P_1+|g_{32}|^2P_2+J\rho}{\sigma^2}\bigg),
    \label{G_d2d:outer:R2R3}\\
R_1+R_2+R_3 &\le \mC\left(\frac{|g_{31}|^2P_1+|g_{32}|^2P_2+J\rho}{\sigma^2}\right)+\mC\left(\frac{(1-\rho^2)|g_{21}|^2P_1}{\sigma^2+(1-\rho^2)|g_{31}|^2P_1}\right).
    \label{G_d2d:outer:R1R2R3}
\end{align}
\end{subequations}
\hrulefill
\end{figure*}
\setcounter{equation}{40}

\subsection{Converse for Scalar Gaussian Model}

We now compute the outer bounds for the scalar Gaussian FD cellular network. The following proposition shows an evaluation of the outer bound in Theorem \ref{theorem:DM:outer} for the case without D2D.
\begin{proposition}
\label{prop:G:outer}
For the Gaussian FD cellular network without D2D, any achievable rate pair $(R_1,R_2)$ must satisfy
\begin{subequations}
\label{G:outer}
\begin{align}
R_1  &\le \mathsf C\bigg(\frac{(1-\rho^2)|g_{21}|^2P_1}{\sigma^2}\bigg),
    \label{G:outer_R1}\\
R_2  &\le \mathsf C\bigg(\frac{(1-\rho^2)|g_{32}|^2P_2}{\sigma^2}\bigg),
    \label{G:outer_R2}\\
R_1+R_2  &\le \mC\left(\frac{|g_{31}|^2P_1+|g_{32}|^2P_2+J\rho}{\sigma^2}\right)\notag\\
&\quad\,+\mC\left(\frac{(1-\rho^2)|g_{21}|^2P_1}{\sigma^2+(1-\rho^2)|g_{31}|^2P_1}\right),
    \label{G:outer_R1R2}
\end{align}
\end{subequations}
for some parameter $\rho\in[-1,1]$, where $J$ is defined in (\ref{def_J}).
\end{proposition}
\begin{IEEEproof}
Let $\rho = \frac{1}{\sqrt{P_1P_2}}\mathbb E\big[X_{1}X_{2}\big]$ and observe that the correlation coefficient $\rho\in[-1,1]$.
We first evaluate the upper bound (\ref{DM:outer_R1}) on $R_1$:
\begingroup
\allowdisplaybreaks
\begin{align}
\label{G:converse_R1}
R_1 &\le I(X_1;Y_2|X_2)\notag\\
&= h(g_{21}X_1+Z_2|X_2) - h(Z_2)\notag\\
&\le  \log_2\Big(\frac{1}{\sigma^2}\mathsf{Var}(g_{21}X_1|X_2)\Big)\notag\\
&\overset{(a)}\le \mathsf C\bigg(\frac{1}{\sigma^2}\bigg(\mathbb E\big[|g_{21}X_1|^2\big]-\frac{\mathbb E^2\big[|g_{21}X_1X_2|\big]}{\mathbb E \big[|X_2|^2\big]}\bigg)\bigg)\notag\\
&=\mathsf C\bigg(\frac{(1-\rho^2)|g_{21}|^2P_1}{\sigma^2}\bigg).
\end{align}
\endgroup
The proof of step $(a)$ in (\ref{G:converse_R1}) is as follows.
{Suppose that we wish to estimate an unknown quantity $S=g_{21}X_1$
based on the observation $Z=X_2$ under a minimum mean squared-error (MMSE) criteria.
The optimal MMSE estimator is $\hat S=\mathbb E\big[S|Z\big]$, and the
minimum MMSE equals to $\mathsf{Var}(g_{21}X_1|X_2)$. If we further restrict
the estimator of $S$ to be a linear function of $Z$, then the corresponding
linear MMSE must be greater than or equal to $\mathsf{Var}(g_{21}X_1|X_2)$.
Since the linear MMSE can be computed analytically as
\begin{align}
\mathsf{LMMSE} &= \mathsf{Var}(S)-\mathsf{Cov}(S,Z)\cdot\mathsf{Var}(Z)^{-1}\cdot\mathsf{Cov}(S,Z)^H\notag\\
&=\mathbb E\big[|g_{21}X_1|^2\big]-\frac{\mathbb E^2\big[|g_{21}X_1X_2|\big]}{\mathbb E \big[|X_2|^2\big]}.
\end{align}
Thus, step $(a)$ in (\ref{G:converse_R1}) is established by using the above $\mathsf{LMMSE}$ as an upper bound on $\mathsf{Var}(g_{21}X_1|X_2)$.}

Likewise, let $S=Y_2$ and $Z=(X_2,Y_3)$. The linear MMSE of this case is
\begin{multline}
\mathsf{LMMSE} = \mathsf{Var}(Y_2) - \mathsf{Cov}(Y_2,[X_2,Y_3])\cdot\mathsf{Var}^{-1}([X_2,Y_3])\\
\cdot\mathsf{Cov}^{H}(Y_2,[X_2,Y_3]),
\end{multline}
with which we can evaluate (\ref{DM:outer_R1R2}) as follows:
\begin{align}
R_1+R_2 &\le I(X_1;Y_2,Y_3|X_2)+I(X_2;Y_3)\notag\\
&= h(Y_2|Y_3,X_2) - h(Z_2,Z_3) + h(Y_3)\notag\\
&\le \log_2\Big(\frac{1}{\sigma^2}\mV(Y_2|Y_3,X_2)\Big) + \log_2\Big(\frac{1}{\sigma^2} \mV(Y_3)\Big)\notag\\
&\overset{(b)}{\le}\log_2\bigg(\frac{\mathsf{LMMSE}}{\sigma^2}\bigg) + \log_2\Big(\frac{1}{\sigma^2} \mV(Y_3)\Big)\notag\\
&\overset{}{=}\mC\left(\frac{(1-\rho^2)|g_{21}|^2P_1}{\sigma^2(1-\rho^2)|g_{31}|^2P_1}\right)\notag\\
&\quad\;+\mC\left(\frac{|g_{31}|^2P_1+|g_{32}|^2P_2+J\rho}{\sigma^2}\right),
    \label{OB:Gaussian_R1R2_proof}
\end{align}
where $(b)$ follows by the aforementioned property of MMSE. The upper bound on $R_2$ can be obtained similarly. 
\end{IEEEproof}


The following proposition evaluates the outer bound (\ref{DM:outer}) for the scalar Gaussian FD cellular network with D2D.

\begin{proposition}
\label{prop:G_d2d:outer}
For the scalar Gaussian FD cellular network with D2D, any achievable rate triple $(R_1,R_2,R_3)$ must satisfy (\ref{G_d2d:outer}) as displayed at the top of the page, for some parameters $\alpha,\beta\in[0,1]$ and $\rho\in[-1,1]$, where $J$ is defined in (\ref{def_J}). 
\end{proposition}
\begin{IEEEproof}
Please see Appendix \ref{proof:G_d2d:outer}.
\end{IEEEproof}

By setting $\alpha=0$ and $\beta=1$ in (\ref{G_d2d:outer:R1}), $\alpha=1$ and $\beta=0$ in (\ref{G_d2d:outer:R2}), and $\rho=1$ throughout, also ignoring the bound (\ref{G_d2d:outer:R3}) on $R_3$, we obtain a simpler outer bound:
{\begin{corollary}
\label{corollary:G_d2d:relaxed_OB}
For the scalar Gaussian FD cellular network with D2D, any achievable rate triple $(R_1,R_2,R_3)$ must satisfy
\allowdisplaybreaks
\setcounter{equation}{46}
\begin{subequations}
\label{G_d2d:outer:gap}
\begin{align}
R_1 &\le \mC\bigg(\frac{|g_{21}|^2P_1}{\sigma^2}\bigg),
    \label{G_d2d:outer:R1:gap}\\
R_2 &\le \mC\bigg(\frac{|g_{32}|^2P_2}{\sigma^2}\bigg),
    \label{G_d2d:outer:R2:gap}\\
R_1+R_3 &\le \mC\bigg(\frac{(|g_{21}|^2+|g_{31}|^2)P_1}{\sigma^2}\bigg),
    \label{G_d2d:outer:R1R3:gap}\\
R_2+R_3 &\le \mC\bigg(\frac{|g_{31}|^2P_1+|g_{32}|^2P_2+J}{\sigma^2}\bigg),
    \label{G_d2d:outer:R2R3:gap}\\
R_1+R_2+R_3 &\le \mC\left(\frac{|g_{31}|^2P_1+|g_{32}|^2P_2+J}{\sigma^2}\right)\notag\\
&\quad+\mC\left(\frac{|g_{21}|^2P_1}{\sigma^2+|g_{31}|^2P_1}\right),
    \label{G_d2d:outer:R1R2R3:gap}
\end{align}
\end{subequations}
where $J$ is defined in (\ref{def_J}).
\end{corollary}
}

We see in the next section that the above simpler outer bound suffices to establish
an approximate capacity region of the scalar Gaussian FD cellular network.

\subsection{Capacity Region to Within One Bit}
\label{subsec:gap}

First we define the notion of the capacity region to within a constant gap.
{\begin{definition}
An achievable rate region $\mathcal R\subseteq\mathbb R^k_+$ is within a constant gap $\delta\ge0$ of the capacity region $\mathcal C$ if $\big((C_1-\delta)^+,\ldots,(C_k-\delta)^+\big)\in\mathcal R$ for any $(C_1,\ldots,C_k)\in\mathcal C$, where $(\cdot)^+=\max\{\cdot,0\}$.
\end{definition}}

\begin{theorem}
\label{theorem:G_d2d:gap}
For the scalar Gaussian FD cellular network with D2D,
the achievable rate region of Proposition \ref{prop:G_d2d:inner1} is within 1 bit of the capacity region under the condition $|g_{31}|\ge|g_{21}|$; the achievable rate region of Proposition \ref{prop:G_d2d:inner2} is within 1 bit of the capacity region under the condition $|g_{31}|<|g_{21}|$. Hence, the achievable rate region of Theorem \ref{theorem:DM_d2d:inner} is within 1 bit of the capacity region.
\end{theorem}
\begin{IEEEproof}
The key is to set the parameters of the inner bounds properly. We set $a=d=0$, $b=1-c$, $e=1$, and $c=\min\{1,\sigma^2/(|g_{21}|^2P_1)\}$ in Proposition \ref{prop:G_d2d:inner1}, and set $a=d=0$, $b=1-c$, $e=1$, and $c=\min\{1,\sigma^2/(|g_{31}|^2P_1)\}$ in Proposition \ref{prop:G_d2d:inner2}. The details are provided in Appendix \ref{appendix:gap}.
\end{IEEEproof}

\begin{remark}
The constant-gap optimality stated in the above theorem can be carried over to the \emph{partially cooperative relay broadcast channel} of \cite{LiangTIT2007a,LiangTIT2007b} since it is a special case of our channel model when $R_2=0$.
\end{remark}

{\begin{remark}
In proving the constant-gap optimality, the newly introduced upper bound on $R_1+R_2+R_3$ plays a key role, whereas the cut-set bound used in \cite{Tannious07} is not tight enough to determine the approximate capacity. We demonstrate this point numerically in Fig.~\ref{fig:GDoF} in Section \ref{sec:simulation}.
\end{remark}}

The use of different rate-splitting strategies depending on the channel condition is
crucial in proving the above result; using either of the two strategies alone
does give a bounded gap to the capacity region. This is because suppose the D2D
channel $g_{31}$ is much stronger than the uplink channel $g_{21}$, we would
want to let node 3 decode the entire $m_1$ so it will benefit from interference
cancellation; in this case, $m_1$ ought not to be split. Likewise, we would not
want to split $m_3$ if $g_{21}$ is much stronger. This is why we should apply
the D2D rate splitting strategy if $|g_{31}|\ge |g_{21}|$ and the uplink rate
splitting strategy otherwise. This approach turns out to be approximately optimal.


We now specialize the result to the without D2D case.

\begin{theorem}
\label{theorem:G:capcity}
For the scalar Gaussian FD cellular network without D2D, the achievable rate region (\ref{G:inner}) of
Proposition \ref{prop:G:inner} is within 1 bit of the capacity region.
\end{theorem}
\begin{IEEEproof}
The achievable rate region (\ref{G:inner}) in Proposition \ref{prop:G:inner} is
equivalent to (\ref{G_d2d:inner2}) when $R_3=0$ (see Remark
\ref{remark:uplink_rate_splitting_reduce_to_without_D2D}), which is itself
obtained by uplink rate splitting and is within a constant gap of 1 bit of the
capacity region when $|g_{31}|<|g_{21}|$. So, when $|g_{31}|<|g_{21}|$, (\ref{G:inner})
must be within 1 bit of the capacity region for the case without D2D.

When $|g_{31}| \ge |g_{21}|$, (\ref{G_d2d:inner1}) achieves the capacity region to
within 1 bit using the D2D rate splitting strategy. But when $R_3=0$,
there is no D2D rate to split. In fact, (\ref{G_d2d:inner1}) reduces to treating
interference as noise, so (\ref{G:inner}) is in fact larger than (\ref{G_d2d:inner1}).
Thus, when $|g_{31}| \ge |g_{21}|$, (\ref{G:inner}) must also be within 1 bit of
the capacity region for the case without D2D.
\end{IEEEproof}

\begin{remark}
It turns out that for the without D2D case, it is actually possible to achieve
the capacity region to within a constant gap using a simpler strategy without
even using the BS as a relay.  However, when there is D2D transmission,
relaying plays a crucial role in enhancing $R_3$ and is necessary for achieving
the capacity region to within a constant gap.
\end{remark}

Furthermore, we show that the value of the constant gap $\delta$ can be further reduced for the without D2D case under a strong interference condition.
\begin{theorem}
\label{prop:G_strong:capacity}
For the scalar Gaussian FD cellular network without D2D, in the \emph{strong interference} regime, defined as the regime in which $|g_{31}|\ge |g_{21}|$, the achievable rate region defined by the set of inequalities in Proposition \ref{prop:G:inner} is within $\frac{1}{2}+\frac{1}{2}\log_2(\frac{\sqrt{2}+1}{2})\approx0.6358$ bits of the capacity region.
\end{theorem}
\begin{IEEEproof}
For each $\rho\in[-1,1]$ in (\ref{G:outer}), we correspondingly let $a=d=\rho^2$, $b=e=1-\rho^2$, and $c=0$ in (\ref{G:inner}). Contrasting the resulting achievable rate region (\ref{G:inner}) with the converse bound (\ref{G:outer}), we find that the gap is determined by the inner and outer bounds of $R_1+R_2$, namely (\ref{G:inner_R1R2}) and (\ref{G:outer_R1R2}). Consequently, using the condition $|g_{31}|\ge |g_{21}|$, we obtain the following upper bound on the gap $\delta$ to the capacity region:
\begin{align}
2\delta &\le 1+ \mathsf C\left(\frac{|g_{31}|^2P_1+|g_{32}|^2P_2+J\rho}{\sigma^2}\right)\notag\\
&\quad\,-\mathsf C\left(\frac{|g_{31}|^2P_1+|g_{32}|^2P_2+J\rho^2}{\sigma^2}\right) \notag \\
&=1+\log_2\left(\frac{\lambda+\rho}{\lambda+\rho^2}\right), \label{gap:2}
\end{align}
where
\begin{equation}
\lambda = \frac{|g_{31}|^2P_1+|g_{32}|^2P_2+\sigma^2}{J}.
\end{equation}
By (\ref{def_J}), it is clear that $\lambda\ge1$. Thus, the solution to the following optimization problem is an upper bound on $\delta$:
\begin{subequations}
\begin{align}
\underset{\lambda,\,\rho}{\text{maximize}}\quad&
\frac{\lambda+\rho}{\lambda+\rho^2}\\
\text{subject to}\quad& \lambda\ge1,\\
&-1\le\rho\le1.
\end{align}
\end{subequations}
This problem is quasi-convex and thus can be optimally solved by considering its first-order condition,
which give rise to the optimal $\lambda^\star=1$ and $\rho^\star=\sqrt{2}-1$. Substituting
$(\lambda^\star,\rho^\star)$ back in (\ref{gap:2}) establishes the theorem.
\end{IEEEproof}

\setcounter{equation}{54}
\begin{figure*}
\begin{equation}
\label{VG_d2d:Psi}
\bm\Psi = \big(\bG_{31}\bK_d+\bG_{31}\bK_c\bG_{31}^H\bQ^H\big)\big(\bK_d+\bQ\bG_{31}\bK_c\bG_{31}^H\bQ^H\big)^{-1}
\big(\bG_{31}\bK_d+\bG_{31}\bK_c\bG_{31}^H\bQ^H\big)^H.
\end{equation}
\hrulefill
\end{figure*}

\section{Vector Gaussian Channel Model}
\label{sec:vector_G}

{The achievable rate region of the FD cellular network can be significantly
enlarged by spatial multiplexing and Marton's broadcast coding if the BS and
the uplink and downlink users are equipped with multiple antennas.
For example, by adding just one more antenna to node 1, it is already possible to
transmit the D2D message and the uplink message in orthogonal spatial
dimensions, thereby achieving a degree-of-freedom (DoF) gain.
Furthermore, while splitting either the uplink message or the D2D message
alone already suffices to achieve the approximate capacity of the scalar
Gaussian case, this is no longer true for the vector Gaussian case.}

This section aims to extend the previous results of the scalar Gaussian channel to
the vector Gaussian channel case.
For the vector Gaussian case, the main challenge in evaluating its achievable
rate region is to decide how to set the auxiliary random variables in
Marton's broadcast coding scheme optimally in Theorem \ref{theorem:DM_d2d:inner}. In the what follows, we provide three possible achievable rate regions all based on the
dirty-paper coding. It is likely that none of these is optimal, but they give
achievable rates that can be easily evaluated and implemented using beamforming
and dirty-paper coding. Here, $m_{1}$ is split into $(m_{10},m_{11})$ and
$m_3$ split into $(m_{30},m_{33})$.  The three achievable rate regions have the same
form as the inner bound in (\ref{DM_d2d:inner}), but differ in what is treated
as dirt and what is treated as noise, and so differ in the values of
$(\mu_1,\ldots,\mu_7)$ in (\ref{Marton:mutual_info}).

\subsubsection{Treating $m_{33}$ as Dirt}
We treat $m_{33}$ as the dirt in the encoding of $m_{11}$ so that node 2 can decode $m_{11}$ as if the interference from $m_{33}$ does not exist; the uplink transmission has priority in this scheme. The resulting mutual information terms in (\ref{Marton:mutual_info}) are computed as
\setcounter{equation}{50}
\begin{subequations}
\label{DPC1}
\begin{align}
\mu_1 &= \log\frac{\big|\bK_c+\bQ\bG_{21}\bK_d\bG_{21}^H\bQ^H\big|}{\big|\bK_c\big|},\\
\mu_2 &= \log\big|\bI+1/\sigma^2\cdot\bG_{31}\bK_c\bG_{21}^H\big|+\mu_1,\\
\mu_3 &= \log\frac{\big|\sigma^2\bI+\bG_{21}(\bK_b+\bK_c+\bK_d)\bG_{21}^H\big|}{\big|\sigma^2\bI+\bG_{21}(\bK_c+\bK_d)\bG_{21}^H\big|}+\mu_2,\\
\mu_4 &= \log\frac{\big|\sigma^2\bI+\bG_{31}(\bK_c+\bK_d)\bG_{31}^H\big|}{\big|\sigma^2\bI+\bG_{31}\bK_c\bG_{31}^H\big|},\\
\mu_5 &= \log\frac{\big|\sigma^2\bI+\bG_{31}\bK_c\bG_{31}^H+\bG_{32}\bK_f\bG_{32}^H\big|}{\big|\sigma^2\bI+\bG_{31}\bK_c\bG_{31}^H\big|},\\
\mu_6 &= \log\frac{\big|\sigma^2\bI+\bG_{31}(\bK_c+\bK_d)\bG^H_{31}+\bG_{32}\bK_f\bG_{32}^H\big|}{\big|\sigma^2\bI+\bG_{31}\bK_c\bG_{31}^H\big|},\\
\mu_7 &= \log\frac{\big|\sigma^2\bI+\bm\Phi\big|}{\big|\sigma^2\bI+\bG_{31}\bK_c\bG_{31}^H\big|},
\end{align}
\end{subequations}
for some $\Lmd_a\in\mathbb C^{L^+_1\times \min\{L^+_1,L^+_2\}}$, $\Lmd_b\in\mathbb C^{L^+_1\times L^+_1}$, $\Lmd_c\in\mathbb C^{L^+_1\times L^+_1}$, $\Lmd_d\in\mathbb C^{L^+_1\times L^+_1}$,$\Lmd_e\in\mathbb C^{L^+_2\times \min\{L^+_1,L^+_2\}}$, and $\Lmd_f\in\mathbb C^{L^+_2\times L^+_2}$ under the power constraints $\mathrm{tr}\big(\bK_{a}+\bK_{b}+\bK_{c}+\bK_{d}\big)\le P_1$ and $\mathrm{tr}\big(\bK_{e}+\bK_{f}\big)\le P_2$, with
\begin{equation}
\bQ = \bK_{c}\bG^H_{21}\big(\sigma^2\bI+\bG_{21}\bK_{c}\bG_{21}^H\big)^{-1},
\end{equation}
and
\begin{multline}
\label{VG_d2d:Phi}
\bm\Phi = \bG_{31}(\bK_{a}+\bK_{b}+\bK_{c}+\bK_{d})\bG^H_{31}+\bG_{32}(\bK_{e}+\bK_{f})\bG^H_{32}\\
+\bG_{31}\bPsi_a\bPsi_e^H\bG^H_{32}+
\bG_{32}\bPsi_e\bPsi_a^H\bG^H_{31},
\end{multline}
where $\bK_{i}=\Lmd_i\Lmd_i^H$, $\forall i\in\{a,b,c,d,e,f\}$. {Node 1 uses $\bm\Lambda_a$, $\bm\Lambda_b$, $\bm\Lambda_c$, and $\bm\Lambda_d$ as the beamformers for the relay message $(m^{t-1}_{10},m^{t-1}_{10})$, the common message $(m^t_{10},m^t_{10})$, the uplink private message $m^t_{11}$, and the D2D private message $m^t_{33}$, respectively, while node 2 uses $\bm\Lambda_e$ and $\bm\Lambda_f$ as the beamformers for the common message $(m^{t-1}_{10},m^{t-1}_{10})$ and the downlink message $m^t_2$, respectively, where $t\in[1:T]$ represents the block index in the block Markov coding.}

\subsubsection{Treating $m_{11}$ as Dirt and Subtracting $m_2$ First at Node 3}
We now treat $m_{11}$ as the dirt in the encoding of $m_{33}$ so that node 3
can decode $m_{33}$ as if the interference from $m_{11}$ does not exist. We
further assume that Node 3 decodes $m_2$ then subtract it first. Thus, the D2D
transmission has priority in this scheme. Using the same variables
$(\Lmd_a,\Lmd_b,\Lmd_c,\Lmd_d,\Lmd_e,\Lmd_f)$ as in the previous scheme, we
evaluate the mutual information terms in (\ref{Marton:mutual_info}) as
\begingroup
\allowdisplaybreaks
\begin{subequations}
\label{DPC2}
\begin{align}
\mu_1 &= \log\frac{\big|\bK_d+\bQ\bG_{31}\bK_c\bG_{31}^H\bQ^H\big|}{\big|\bK_d\big|},\\
\mu_2 &= \log\frac{\big|\sigma^2\bI+\bG_{31}(\bK_c+\bK_d)\bG_{31}^H\big|}{\big|\sigma^2\bI+\bG_{31}\bK_d\bG_{31}^H\big|},\\
\mu_3 &= \log\frac{\big|\sigma^2\bI+\bG_{31}(\bK_b+\bK_c+\bK_d)\bG_{31}^H\big|}{\big|\sigma^2\bI+\bG_{31}\bK_d\bG_{31}^H\big|},\\
\mu_4 &= \log\big|\bI+1/\sigma^2\cdot\bG_{31}\bK_d\bG_{31}^H\big|+\mu_1,\\
\mu_5 &= \log\frac{\big|\sigma^2\bI+\bG_{31}(\bK_c+\bK_d)\bG_{31}^H+\bG_{32}\bK_f\bG_{32}^H\big|}{\big|\sigma^2\bI+\bG_{31}(\bK_c+\bK_d)\bG_{31}^H\big|}+\mu_4\notag\\
&\quad\,-\log\frac{\big|\sigma^2\bI+\bG_{31}(\bK_c+\bK_d)\bG_{31}^H+\bG_{32}\bK_f\bG_{32}^H\big|}{\big|\sigma^2\bI+\bG_{31}(\bK_c+\bK_d)\bG_{31}^H+\bG_{32}\bK_f\bG_{32}^H-\bm\Psi\big|},\\
\mu_6 &= \log\frac{\big|\sigma^2\bI+\bG_{31}(\bK_c+\bK_d)\bG_{31}^H+\bG_{32}\bK_f\bG_{32}^H\big|}{\big|\sigma^2\bI+\bG_{31}(\bK_c+\bK_d)\bG_{31}^H\big|}+\mu_4,\notag\\
\mu_7 &= \log\frac{\big|\sigma^2\bI+\bm\Phi\big|}{\big|\sigma^2\bI+\bG_{31}(\bK_c+\bK_d)\bG_{31}^H\big|}+\mu_4-\mu_1,
\end{align}
\end{subequations}
\endgroup
where a new variable $\bm\Psi$ is defined in (\ref{VG_d2d:Psi}) as displayed at the top of the page.

\subsubsection{Treating $m_{11}$ as Dirt and Treating $m_2$ as Noise at Node 3}
We still treat $m_{11}$ as the dirt but treat $m_2$ as the noise in the encoding of $m_{33}$, so the downlink transmission has priority in this scheme. The mutual information terms in (\ref{Marton:mutual_info}) then become
\setcounter{equation}{55}
\begin{subequations}
\label{DPC3}
\begin{align}
\mu_1 &= \log\frac{\big|\bK_d+\bQ\bG_{31}\bK_c\bG_{31}^H\bQ^H\big|}{\big|\bK_d\big|},\\
\mu_2 &= \log\frac{\big|\sigma^2\bI+\bG_{31}(\bK_c+\bK_d)\bG_{31}^H\big|}{\big|\sigma^2\bI+\bG_{31}\bK_d\bG_{31}^H\big|},\\
\mu_3 &= \log\frac{\big|\sigma^2\bI+\bG_{31}(\bK_b+\bK_c+\bK_d)\bG_{31}^H\big|}{\big|\sigma^2\bI+\bG_{31}\bK_d\bG_{31}^H\big|},\\
\mu_4 &= \log\frac{\big|\sigma^2\bI+\bG_{31}(\bK_c+\bK_d)\bG_{31}^H\big|}{\big|\sigma^2\bI+\bG_{31}(\bK_c+\bK_d)\bG_{31}^H-\bm\Psi\big|},\\
\mu_5 &= \log\frac{\big|\sigma^2\bI+\bG_{31}(\bK_c+\bK_d)\bG_{31}^H+\bG_{32}\bK_f\bG_{32}^H\big|}{\big|\sigma^2\bI+\bG_{31}(\bK_c+\bK_d)\bG_{31}^H\big|}+\mu_4\notag\\
&\quad-\log\frac{\big|\sigma^2\bI+\bG_{31}\bK_d\bG_{31}^H+\bG_{32}\bK_f\bG_{32}^H\big|}{\big|\sigma^2\bI+\bG_{32}\bK_f\bG_{32}^H\big|}-\mu_1,\\
\mu_6 &= \log\frac{\big|\sigma^2\bI+\bG_{31}(\bK_c+\bK_d)\bG_{31}^H+\bG_{32}\bK_f\bG_{32}^H\big|}{\big|\sigma^2\bI+\bG_{31}(\bK_c+\bK_d)\bG_{31}^H\big|}+\mu_4,\notag\\
\mu_7 &= \log\frac{\big|\sigma^2\bI+\bm\Phi\big|}{\big|\sigma^2\bI+\bG_{31}(\bK_c+\bK_d)\bG_{31}^H\big|}+\mu_4-\mu_1,
\end{align}
\end{subequations}
where the variables are all defined as before.

{It is challenging to prove the constant-gap optimality for the above achievable
rate regions. 
But, in the without D2D case, treating $m_{33}$ as dirt method
reduces to a vector generalization of Proposition \ref{prop:G:inner}, 
and it can be
shown that the resulting achievable rate region reaches the capacity to within
$\max\big\{\min\{L^+_1,L^-_2\},\min\{L^+_2,L^-_3\},\frac{1}{2}
\min\{L^+_1,L^-_3\}\big\}$ bits for the vector Gaussian channel model without D2D
by setting $\bK_d=\sigma^2\big(\frac{\sigma^2}{ P_1}\bI+\bG^H_{31}\bG_{31}\big)^{-1}$, $\bK_c=P_1\bI-\bm\Sigma_c$, and $\bK_e=\sqrt{P_2}\bI$, along with $(\bPsi_a,\bPsi_c,\bPsi_e)$ being zero matrices.}

\begin{figure*}[t]
\begin{minipage}[t]{0.33\linewidth}
\psfrag{I}[r][t]{\footnotesize $\mathsf{INR}$ (dB)}
\psfrag{S}[l][t]{\footnotesize $\mathsf{SNR}$ (dB)}
\psfrag{G}[][]{\footnotesize Gap (bit/s/Hz)}
\centerline{\includegraphics[width=6.7cm]{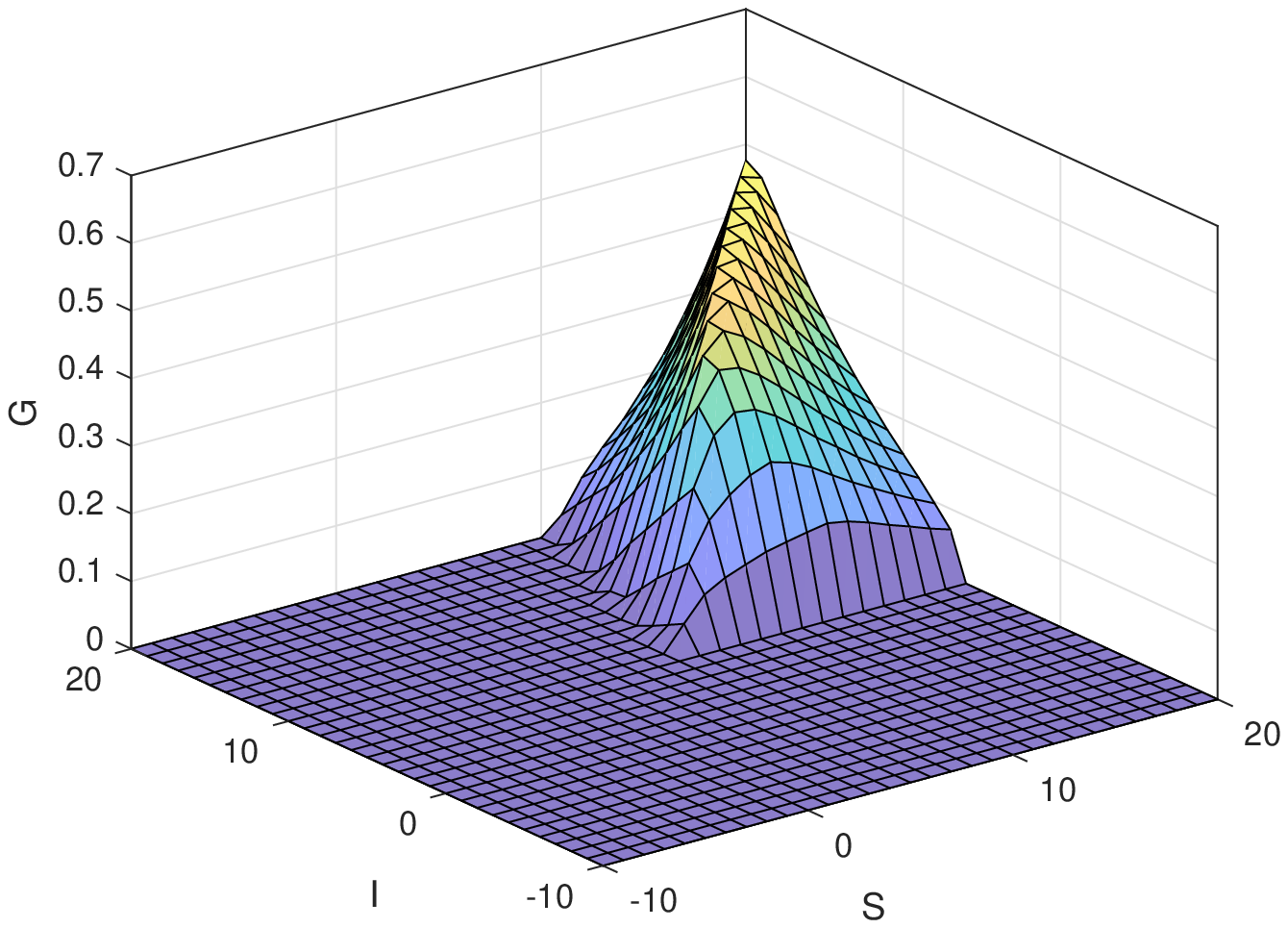}}
\centerline{\footnotesize(a) Downlink stronger than uplink}
\end{minipage}
\begin{minipage}[t]{0.33\linewidth}
\psfrag{I}[r][t]{\footnotesize $\mathsf{INR}$ (dB)}
\psfrag{S}[l][t]{\footnotesize $\mathsf{SNR}$ (dB)}
\psfrag{G}[][]{\footnotesize Gap (bit/s/Hz)}
\centerline{\includegraphics[width=6.7cm]{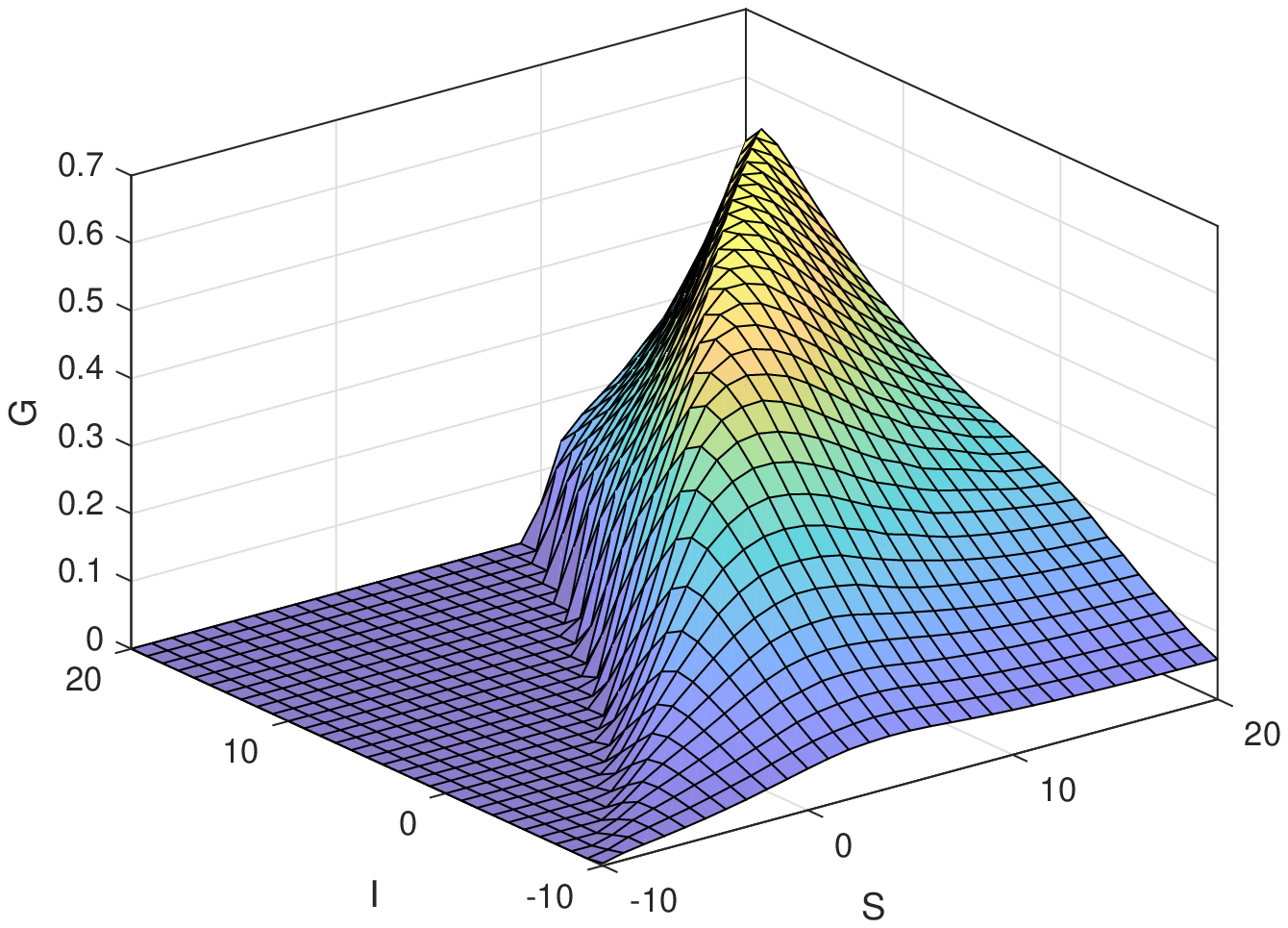}}
\centerline{\footnotesize(b) Downlink and uplink equally strong}
\end{minipage}
\begin{minipage}[t]{0.33\linewidth}
\psfrag{I}[r][t]{\footnotesize $\mathsf{INR}$ (dB)}
\psfrag{S}[l][t]{\footnotesize $\mathsf{SNR}$ (dB)}
\psfrag{G}[][]{\footnotesize Gap (bit/s/Hz)}
\centerline{\includegraphics[width=6.7cm]{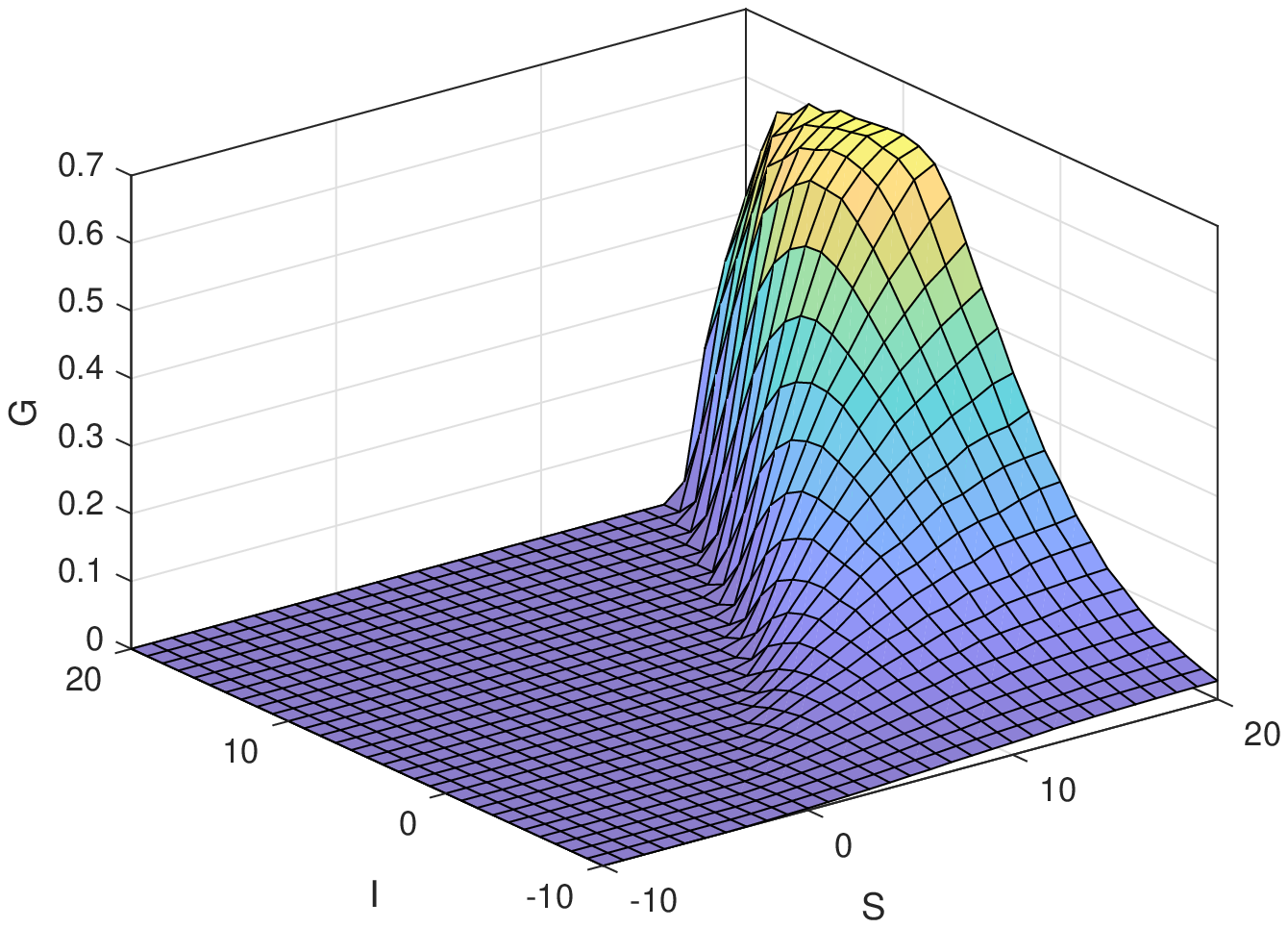}}
\centerline{\footnotesize(c) Uplink stronger than downlink}
\end{minipage}
\caption{{The gap between the achievable rate region (\ref{G:inner}) and the converse (\ref{G:outer}) for the scalar Gaussian FD cellular network without D2D, where the parameters of the bounds are globally optimized by exhaustive search. We consider three different settings: (a) $|g_{32}|^2P_2=5|g_{21}|^2P_1$; (b) $|g_{32}|^2P_2=|g_{21}|^2P_1$; (c)  $5|g_{32}|^2P_2=|g_{21}|^2P_1$.}}
\label{fig:gap}
\end{figure*}

\section{Numerical Examples}
\label{sec:simulation}

To demonstrate the above capacity analyses, we provide some numerical examples of the scalar Gaussian FD cellular network. Throughout this section, define $\mathsf{SNR}=|g_{21}|^2P_1/\sigma^2$ and $\mathsf{INR}=|g_{31}|^2P_1/\sigma^2$.

\subsection{Without D2D Case}

\begin{figure}[t]
\begin{minipage}{1.0\linewidth}
\psfrag{x}[][b]{\footnotesize User-to-User Distance (meter)}
\psfrag{y}[][t]{\footnotesize $\max\min\{R_1,R_2\}$ (bit/s/Hz)}
\centerline{\includegraphics[width=9.7cm]{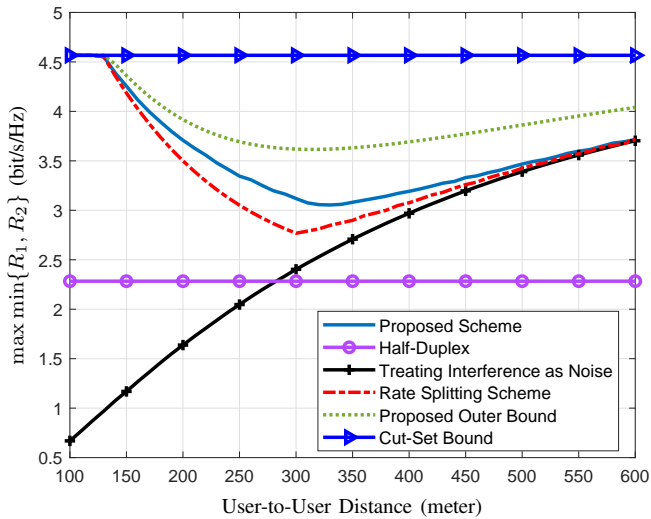}}
\caption{Symmetric rate $\min\{R_1,R_2\}$ in a Gaussian channel without D2D where the user-to-BS distance is 300 meters.}
\label{fig:nod2d}
\end{minipage}
\end{figure}

\begin{figure}[t]
\begin{minipage}{1.0\linewidth}
\psfrag{x}[][b]{\footnotesize User-to-User Distance (meter)}
\psfrag{y}[][t]{\footnotesize $R_1+R_2$ (bit/s/Hz)}
\centerline{\includegraphics[width=9.7cm]{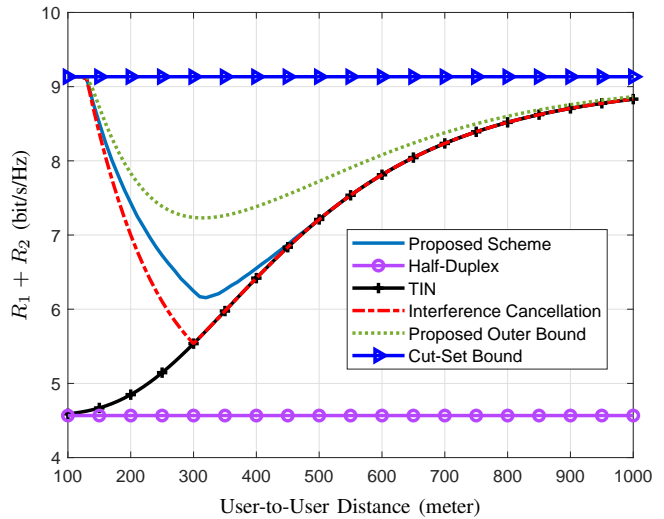}}
\caption{Sum rate $R_1+R_2$ in a Gaussian channel without D2D where the user-to-BS distance is 300 meters.}
\label{fig:sum_rate}
\end{minipage}
\end{figure}

For the scalar Gaussian channel model without D2D,
Theorem \ref{theorem:G:capcity} shows that the gap
$\delta$ between the inner bound (\ref{G:inner}) and the outer bound
(\ref{G:outer}) is at most 1 bit/s/Hz under a particular choice of the
parameters $(a,b,c,d,e,\rho)$. Fig.~\ref{fig:gap} shows the minimum value of
$\delta$ under various channel conditions {with the parameters $(a,b,c,d,e,\rho)$
globally optimized by exhaustive search.} Observe
that the gap is equal to zero if $\mathsf{INR}$ is sufficiently greater than
$\mathsf{SNR}$ because of Theorem \ref{prop:G_very_strong:capacity}. In all cases,
the gap is less than 1 bit as expected.


We further evaluate the data rates in an FD cellular network in the following
topology. The BS is at the center of the cell; the uplink/downlink user-to-BS
distance is fixed at $300$m, the user-to-user distance is set to
different values. Let the maximum transmit power spectral density (PSD) be
$-47$ dBm/Hz for both uplink and downlink; we set the PSD of the background
noise be $-169$ dBm/Hz. The channel magnitude is modeled as
$-128.1-37.6\log_{10}(\mathsf{dist})$ in dB scale, $\mathsf{dist}$ representing the distance (in km). In addition to the inner bound (\ref{G:inner}), {the outer bound (\ref{G:outer}), and the cut-set bound,} we consider the following baseline methods:
\begin{itemize}
    \item Half-duplex with separate uplink and downlink; 
\item Treating interference as noise with FD uplink/dowlink; 
    \item {Rate splitting scheme:} The uplink message is split
for interference cancellation, but without relaying by the BS.
\end{itemize}

\begin{figure*}[t]
\begin{minipage}[t]{0.33\linewidth}
\psfrag{I}[r][t]{\footnotesize $\mathsf{INR}$ (dB)}
\psfrag{S}[l][t]{\footnotesize $\mathsf{SNR}$ (dB)}
\psfrag{G}[][]{\footnotesize Gap (bit/s/Hz)}
\centerline{\includegraphics[width=6.7cm]{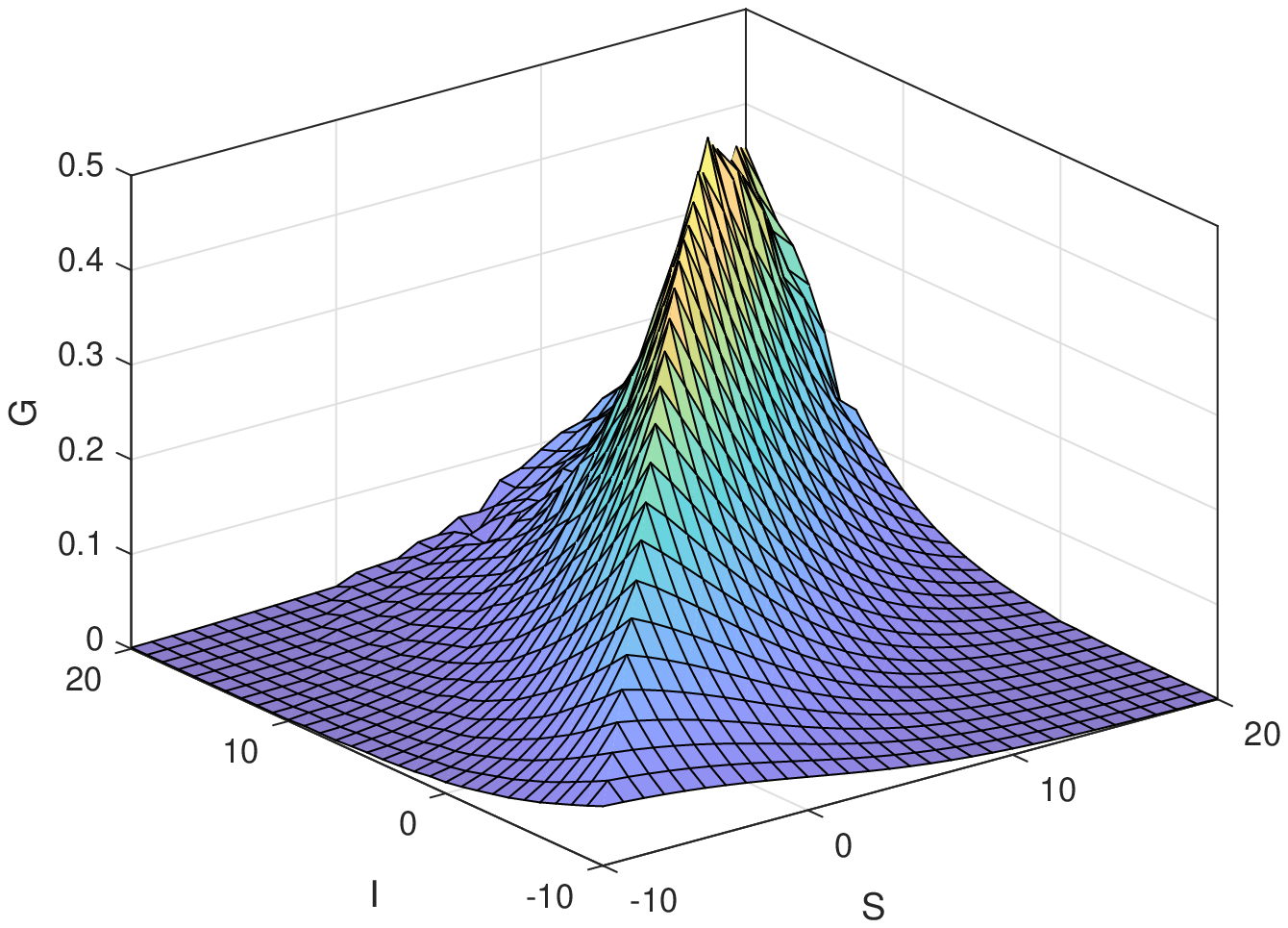}}
\centerline{\footnotesize(a) Downlink stronger than uplink}
\end{minipage}
\begin{minipage}[t]{0.33\linewidth}
\psfrag{I}[r][t]{\footnotesize $\mathsf{INR}$ (dB)}
\psfrag{S}[l][t]{\footnotesize $\mathsf{SNR}$ (dB)}
\psfrag{G}[][]{\footnotesize Gap (bit/s/Hz)}
\centerline{\includegraphics[width=6.7cm]{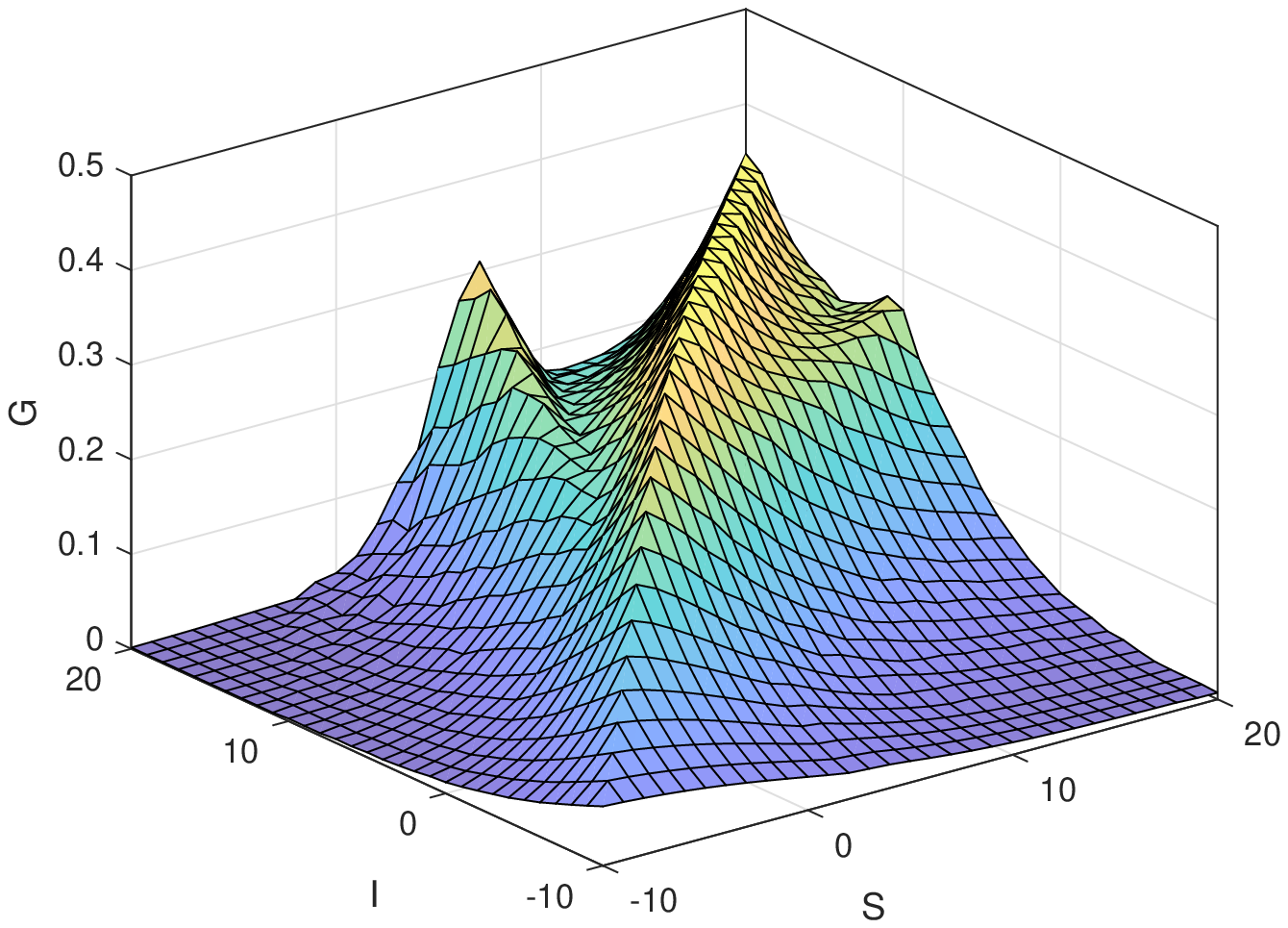}}
\centerline{\footnotesize(b) Downlink and uplink equally strong}
\end{minipage}
\begin{minipage}[t]{0.33\linewidth}
\psfrag{I}[r][t]{\footnotesize $\mathsf{INR}$ (dB)}
\psfrag{S}[l][t]{\footnotesize $\mathsf{SNR}$ (dB)}
\psfrag{G}[][]{\footnotesize Gap (bit/s/Hz)}
\centerline{\includegraphics[width=6.7cm]{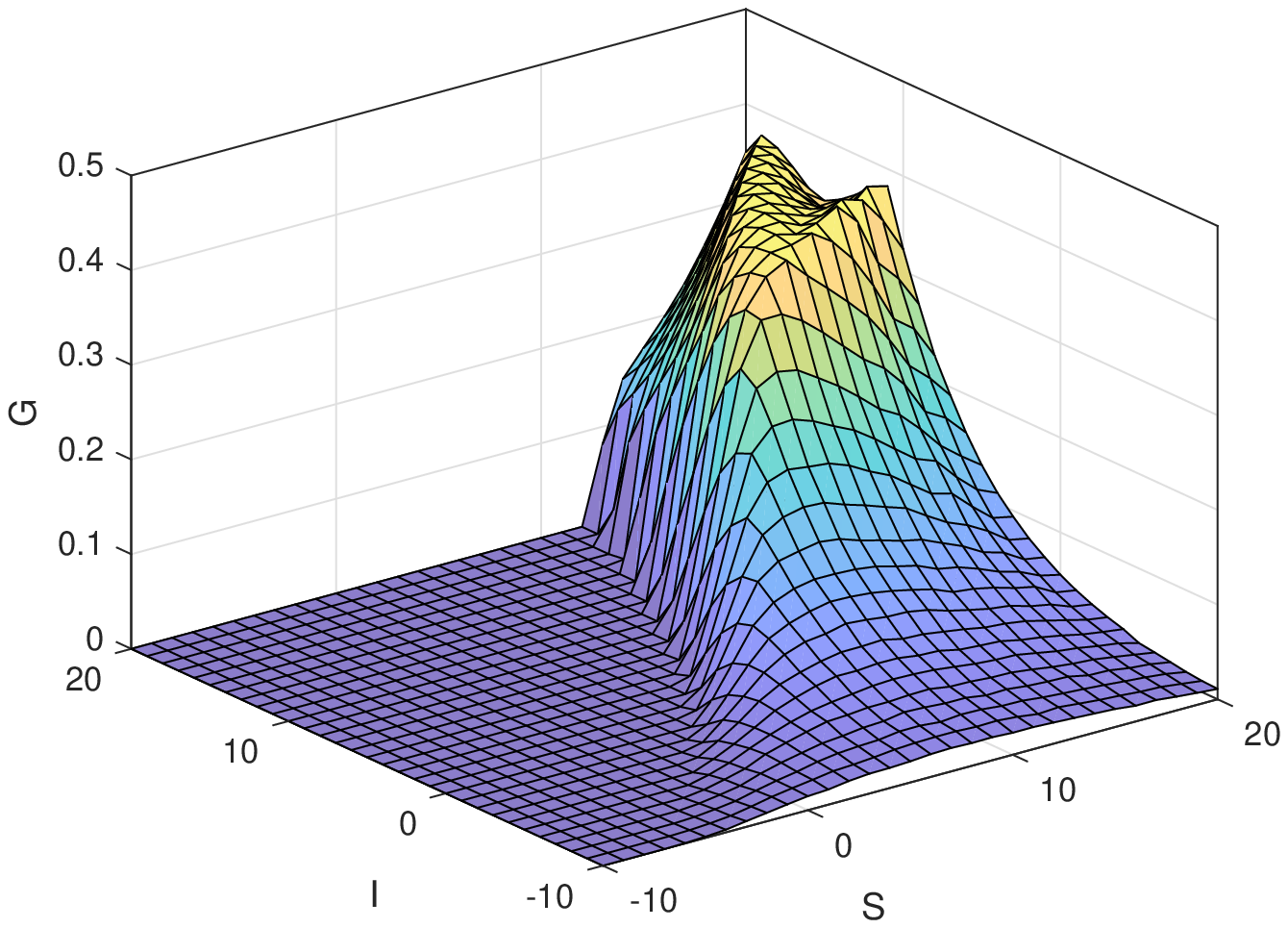}}
\centerline{\footnotesize(c) Uplink stronger than downlink}
\end{minipage}
\caption{{The gap between the achievable rate region, i.e., the union of (\ref{G_d2d:inner1}) and (\ref{G_d2d:inner2}), and the converse (\ref{G_d2d:outer}) for the scalar Gaussian FD cellular network with D2D, where the parameters of the bounds are globally optimized by exhaustive search. We consider three different settings: (a) $|g_{32}|^2P_2=5|g_{21}|^2P_1$; (b) $|g_{32}|^2P_2=|g_{21}|^2P_1$; (c)  $5|g_{32}|^2P_2=|g_{21}|^2P_1$.}}
\label{fig:d2d_gap}
\end{figure*}

Fig.~\ref{fig:nod2d} compares the symmetric uplink-downlink rate
$\min\{R_1,R_2\}$ in the without D2D case. It shows that the proposed BS-aided
scheme is more effective than simple interference cancellation, gaining up to
about $0.5$ bits/s/Hz. 
Observe that the proposed scheme shows the most benefit when the user-to-user distance is not too close or too far.
{Moreover, Fig.~\ref{fig:sum_rate} compares the sum rate of various schemes. It shows that the half-duplex scheme is inferior to all the FD schemes in terms of the total throughput.} 

\subsection{D2D Case}

\begin{figure}[t]
\begin{minipage}[t]{1.0\linewidth}
\psfrag{x}[][b]{\footnotesize $R_3$ (bit/s/Hz)}
\psfrag{y}[][t]{\footnotesize $\min\{R_1,R_2\}$ (bit/s/Hz)}
\centerline{\includegraphics[width=9.7cm]{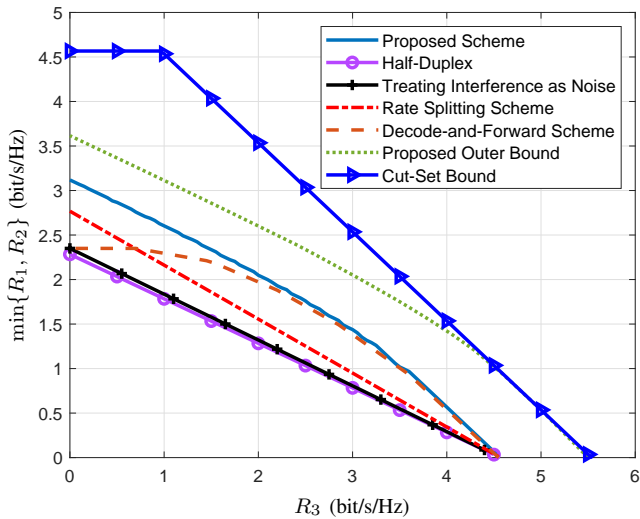}}
\caption{Trade-off between $R_3$ and $\min\{R_1,R_2\}$ in the D2D case.}
\label{fig:d2d}
\end{minipage}
\end{figure}

For the scalar Gaussian channel model with D2D,
we compare the inner bound, i.e., the union of (\ref{G_d2d:inner1}) and
(\ref{G_d2d:inner2}), with the outer bound (\ref{G_d2d:outer}) to find the
minimum gap between them. Again, the parameters of these bounds are optimized
via exhaustive search. The minimum gap with respect to different
$(\mathsf{INR},\mathsf{SNR})$ pairs is plotted in Fig.~\ref{fig:d2d_gap}. 
We again observe that the gap is always less than 1. Moreover,
although
we have not determined the exact capacity region for the D2D case, the plot of
case (c) shows that the symmetric rate achieved by the proposed scheme is close
to the capacity when the cross-channel is sufficiently stronger than the uplink
channel and when the uplink channel is stronger than the downlink channel.

We further evaluate the data rates with D2D under the same setting as
considered for Fig.~\ref{fig:nod2d} except that the user-to-user distance is
now fixed at 300 meters. The three baselines are extended to the D2D case by
time or frequency division multiplex of cellular and D2D traffic. Moreover, we
introduce a new decode-and-forward benchmark
corresponding to the scheme of \cite{Tannious07}. Fig.~\ref{fig:d2d} shows the
trade-off between the D2D rate and the symmetric uplink-downlink rate.
The proposed scheme achieves a larger rate region than the baseline schemes. {Moreover, Fig.~\ref{fig:d2d} shows that at user-to-user distance of 300 meters, the D2D rate $R_3$ can be up to 4.5 bits/s/Hz by using the proposed scheme, whereas the transmission rate from node 1 to node 3 through the BS is only $\min\{R_1,R_2\}=3.2$ bits/s/Hz in the without D2D case as shown in Fig.~\ref{fig:nod2d}.}


\begin{figure}[t]
\psfrag{x}[][b]{\small $\kappa$}
\psfrag{y}[][t]{\small $d_\text{sym}$}
\begin{minipage}[t]{1.0\linewidth}
\centerline{\includegraphics[width=9.7cm]{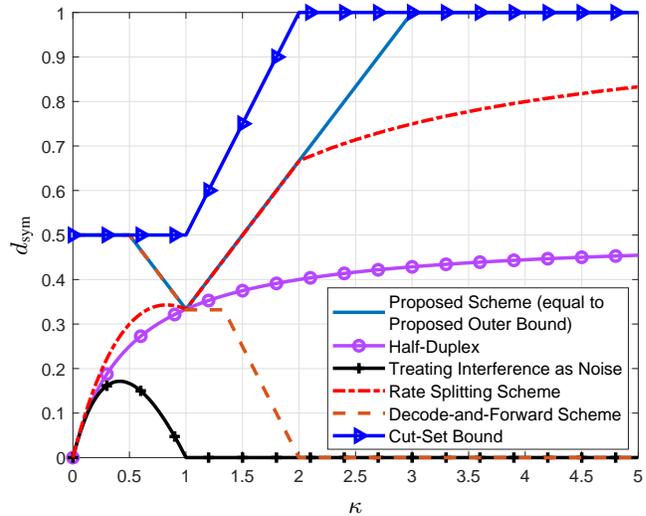}}
\caption{Symmetric GDoF $d_{\text{sym}}$ versus $\kappa=\frac{\log\mathsf{INR}}{\log\mathsf{SNR}}$ in the D2D case.}
\label{fig:GDoF}
\end{minipage}
\end{figure}

{Finally, we compare the asymptotic behavior of the different schemes in terms
of the \emph{generalized degree of freedom} (GDoF) \cite{etkin2008gap}.
We restrict to the channel parameters for which $|g_{21}|P_1=|g_{32}|P_2$,
then let both $\mathsf{INR}$ and $\mathsf{SNR}$ go to infinity, while keeping
their ratio in the log-domain constant, i.e., fixing
$\kappa=\frac{\log\mathsf{INR}}{\log\mathsf{SNR}}$. The symmetric GDoF is
defined as
\begin{equation}
d_\text{sym}\triangleq\lim_{\mathsf{SNR}\rightarrow\infty\atop{\mathsf{INR}\rightarrow\infty}}\frac{\min\{R_1,R_2,R_3\}}{\log \mathsf{SNR}}.
\end{equation}
This definition is motivated by regarding the FD cellular channel model as a
Z-interference channel that has a one-way interference from uplink to downlink.
Here, $\mathsf{SNR}$ and $\mathsf{INR}$ both tend to infinity, while
$\mathsf{INR}=\mathsf{SNR}^{\kappa}$ for some fixed $\kappa\in[0,\infty)$.}

Fig.~\ref{fig:GDoF} shows the GDoF achieved by the different
schemes. {The GDoF of the proposed scheme is optimal
since the proposed scheme attains the capacity of scalar Gaussian FD
network to within a constant gap. Now, because there exists a gap
between the optimal GDoF curve and the cut-set bound, this means
cut-set bound can be arbitrarily loose.}

Observe that the optimal GDoF curve consists of four
segments, i.e., $[0,0.5)$, $[0.5,1)$, $[1,3)$, and $[3,\infty)$ with respect to
$\kappa$. When $\kappa\in[0,0.5)$, because the channel between node 1 and node
3 is very weak, it is useless in terms of GDoF, thus the optimal GDoF does not
change with $\kappa$ in this interval. When $\kappa\in[0.5,1)$, the uplink rate
splitting scheme of Proposition \ref{prop:G_d2d:inner2} attains the optimal
GDoF. When $\kappa\in[1,3)$, the D2D rate splitting scheme of Proposition
\ref{prop:G_d2d:inner1} attains the optimal GDoF. Furthermore, when
$\kappa\in[3,\infty)$, the channel between node 1 and node 3 is strong enough
to let node 3 sequentially recover $(m_1,m_2,m_3)$ by successive cancellation
without relaying at the BS. In this situation, node 1 allocates power
$(1-\mathsf{SNR}^{1-\kappa})P_1$ to the transmission of $m_1$ and allocates power
$\mathsf{SNR}^{1-\kappa}P_1$ to the transmission of $m_3$, while node 2 transmits $m_2$ at
the full power $P_2$. After subtracting the self-interference, node 2 removes
$m_3$ and then decodes $m_1$ via the successive cancellation. Node 3 first
decodes $m_1$, then $m_2$, and finally $m_3$ via the successive cancellation.
As a result, each transmission attains a GDoF of 1. Fig.~\ref{fig:GDoF} shows
that all the other schemes are suboptimal in terms of GDoF.

\section{Conclusion}
\label{sec:conclusion}

This paper analyzes the maximum transmission rates of the uplink
message, the downlink message, and the D2D message in a wireless cellular
network with the FD BS and two half-duplex user terminals. This FD cellular
network can be modeled as a relay broadcast channel with side message. We
propose new strategies to enlarge the existing achievable rate regions. A crucial component of the proposed new strategies is to use the BS as a relay to facilitate cancelling
interference. We further provide novel converse results that are strictly
tighter than the cut-set bound and the state-of-the-art outer bound.
For the without D2D case,
our proposed schemes achieve the capacity regions to within a constant gap
for the Gaussian channel case, even when the BS and users are deployed with multiple antennas.
For the D2D case, a constant-gap optimality is
established for the scalar Gaussian channel case. 


\appendices

\IEEEpeerreviewmaketitle

\section{Proof of Theorem \ref{theorem:DM_d2d:outer}}
\label{proof:DM_d2d:outer}

We first verify the upper bound (\ref{DM_d2d:outer:ff}) on $R_1+R_2+R_3$:
\begin{align}
& N(R_1+R_2+R_3-\epsilon_N)\notag\\
&\le I(M_1;\bY^N_2) + I(M_2;\bY^N_3) + I(M_3;\bY^N_3)\notag\\
&\le I(M_1,M_3;\bY^N_2,\bY^N_3) + I(M_2;\bY^N_3)\notag\\
&\overset{(a)}{=} I(M'_1;\bY^N_2,\bY^N_3)+I(M_2;\bY^N_3)\notag\\
&\overset{(b)}{\le} NI(X_1;Y_2,Y_3|X_2) + NI(X_2;Y_3),
\end{align}
where we use $M'_1$ to denote $(M_1,M_3)$ and thus $(a)$ holds; $(b)$ follows by treating $M'_1$ as $M_1$ in (\ref{DM_proof:outer_R1R2}).

We deal with those inequalities having $U$ or $V$ in the rest of the proof. The cut-set bound on $R_1$ is $R_1\le I(X_1;Y_2|X_2)$, but this must be loose because some portion of $X_1$ carrying the D2D message $m_3$ is independent of $m_1$. (We do not have this issue in the without D2D scenario.) Therefore, we introduce an auxiliary variable $U_n$ to represent the part of $X_1$ ``related'' to $R_1$, as follows:
\begin{equation}
U_n = (M_1,M_2,\bY^{n-1}_2,\bY^{n-1}_3),\;n\in[1:N],
\end{equation}
where $M_2$, $\bY^{n-1}_2$, and $\bY^{n-1}_3$ are due to the cut-set bounds on $R_1$ and $R_3$. Note that $M_3$ is excluded from $U_n$. Another auxiliary variable $V_n$ is similarly motivated:
\begin{equation}
V_n = (M_2,M_3,\bY^{n-1}_2,\bY^{n-1}_3),\;n\in[1:N].
\end{equation}

We then apply $U$ and $V$ to improve the cut-set bound. First, consider the bound on $R_1$ with $U_n$:
\begingroup
\allowdisplaybreaks
\begin{align}
N(R_1-\epsilon_N)&\le I(M_1;\bY^N_2,\bX^N_2)\notag\\
&\le I(M_1;\bY^N_2,\bX^N_2|M_2)\notag\\
&\overset{(c)}= I(M_1;\bY^N_2|M_2)\notag\\
&=\sum^N_{n=1}I(M_1;Y_{2n}|M_2,\bY^{n-1}_2)\notag\\
&\overset{(d)}=\sum^N_{n=1}I(M_1;Y_{2n}|M_2,\bY^{n-1}_2,X_{2n})\notag\\
&\le\sum^N_{n=1}I(M_1,M_2,\bY^{n-1}_2,\bY^{n-1}_3;Y_{2n}|X_{2n})\notag\\
&=\sum^N_{n=1}I(U_n;Y_{2n}|X_{2n})\notag\\
&\le NI(U;Y_2|X_2),
\end{align}
\endgroup
where both of $(c)$ and $(d)$ follow since $X_{2n}$ is a function of $(M_2,\bY^{n-1})$. We then use $V_n$ to give another bound on $R_1$:
\begingroup
\allowdisplaybreaks
\begin{align}
N(R_1-\epsilon_N) &\le I(M_1;\bY^N_2,\bX^N_2)\notag\\
&\le I(M_1;\bY^N_2,\bX^N_2|M_2,M_3)\notag\\
&\overset{(e)}= I(M_1;\bY^N_2,\bY^N_3|M_2,M_3)\notag\\
&= \sum^N_{n=1} I(M_1;Y_{2n},Y_{3n}|M_2,M_3,\bY^{n-1}_2,\bY^{n-1}_3)\notag\\
&\overset{(f)}= \sum^N_{n=1} I(X_{1n};Y_{2n},Y_{3n}|M_2,M_3,\bY^{n-1}_2,\bY^{n-1}_3)\notag\\
&\overset{(g)}= \sum^N_{n=1} I(X_{1n};Y_{2n},Y_{3n}|X_{2n},V_n)\notag\\
&\le NI(X_1;Y_2,Y_3|X_2,V),
    \label{DM_d2d_proof:R1+}
\end{align}
\endgroup
where $(e)$ and $(g)$ follow since $X_{2n}$ is a function of $(M_2,\bY^{n-1}_2)$, $(f)$ is due to the facts that $X_{1n}$ is a function of $(M_1,M_3)$ and that $M_1\rightarrow X_{1n}\rightarrow (Y_{2n},Y_{3n})$ form a Markov chain conditioned on $(M_2,M_3,\bY^{n-1}_2,\bY^{n-1}_3)$.

Further, by using $U_n$, we establish an outer bound on $R_3$:
\begingroup
\allowdisplaybreaks
\begin{align}
N(R_3-\epsilon_N) &\le I(M_3;\bY^N_3)\notag\\
&\le I(M_3;\bY^N_2,\bY^N_3|M_1,M_2)\notag\\
&= \sum^N_{n=1} I(M_3;Y_{2n},Y_{3n}|M_1,M_2,\bY^{n-1}_2,\bY^{n-1}_3)\notag\\
&\overset{(h)}= \sum^N_{n=1} I(X_{1n};Y_{2n},Y_{3n}|M_1,M_2,\bY^{n-1}_2,\bY^{n-1}_3)\notag\\
&= \sum^N_{n=1} I(X_{1n};Y_{2n},Y_{3n}|U_n,X_{2n})\notag\\
&\le NI(X_1;Y_2,Y_3|U,X_2),
\end{align}
\endgroup
where $(h)$ follows by the same reason as for step $(f)$ in (\ref{DM_d2d_proof:R1+}). Finally, the auxiliary variable $V_n$ gives rise to the following outer bound on $R_3$:
\begingroup
\allowdisplaybreaks
\begin{align}
N(R_3-\epsilon_N) &\le I(M_3;\bY^N_3)\notag\\
&\le I(M_3;\bY^N_2,\bY^N_3|M_2)\notag\\
&= \sum^N_{n=1}I(M_3;Y_{2n},Y_{3n}|M_2,\bY^{n-1}_2,\bY^{n-1}_3)\notag\\
&= \sum^N_{n=1}I(M_3;Y_{2n},Y_{3n}|M_2,\bY^{n-1}_2,\bY^{n-1}_3,X_{2n})\notag\\
&\le \sum^N_{n=1}I(M_2,M_3,\bY^{n-1}_2,\bY^{n-1}_3;Y_{2n},Y_{3n}|X_{2n})\notag\\
&= \sum^N_{n=1}I(V_n;Y_{2n},Y_{3n}|X_{2n})\notag\\
&\le NI(V;Y_2,Y_3|X_2).
\end{align}
\endgroup
Summarizing the above results gives the outer bound in this theorem.

\section{Proof of Proposition \ref{prop:G_d2d:outer}}
\label{proof:G_d2d:outer}

We first introduce a useful lemma:
\begin{lemma}
\label{lemma:Y'}
Letting
\begin{equation}
Y' = \frac{g_{21}Y_2+g_{31}Y_3}{\sqrt{|g_{21}|^2+|g_{31}|^2}},
\end{equation}
we have
$I(X_1;Y_2,Y_3|U,X_2)=I(X_1;Y'|U,X_2)$.
\end{lemma}
\begin{IEEEproof}
Observe that
\begin{align}
&I(X_1;Y_2,Y_3|U,X_2)\notag\\
&=I(X_1,Y_2,Y'|U,X_2)\notag\\
&=I(X_1;Y'|U,X_2)+I(X_1;Y_2|U,X_2,Y').
\label{Y'_transform}
\end{align}
Also, with the shorthand
\begin{equation}
\label{Z'_2}
Z'_2 = \frac{|g_{31}|^2Z_2-g_{21}g_{31}Z_3}{|g_{21}|^2+|g_{31}|^2},
\end{equation}
we show that
\begin{align}
I(X_1;Y_2|U,X_2,Y')&\overset{(a)}{\le} I(X_1;Y_2|X_2,Y')\notag\\
&= I(X_1;Z'_2|X_2,Y')\notag\\
&\overset{(b)}=0,
\end{align}
where step $(a)$ follows since $U\rightarrow X_1\rightarrow Y_2$ form a Markov chain conditioned on $(X_2,Y')$, step $(b)$ follows since $Z'_2$ is independent of any of $(X_1,X_2,Y')$. By the squeeze theorem, we must have $I(X_1;Y_2|U,X_2,Y')=0$. Substituting this result back in (\ref{Y'_transform}) verifies the lemma.
\end{IEEEproof}

Equipped with the above lemma, we continue to prove Proposition \ref{prop:G_d2d:outer}. We focus on the inequalities (\ref{G_d2d:outer:R1}) and (\ref{G_d2d:outer:R3}) which involve the use of auxiliary variables $U$ and $V$ from Theorem \ref{theorem:DM_d2d:outer}. Again, use $\rho\in[-1,1]$ to denote the correlation coefficient $\frac{1}{\sqrt{P_1P_2}}\mathbb E\big[X_{1}X_{2}\big]$. It can be shown that
\begin{align}
&\log_2(2\pi e\sigma^2) \notag\\
&\le h(Y'|U,X_1,X_2)\notag\\
&\le h(Y'|U,X_2)\notag\\
&\le h(Y'|X_2)\notag\\
&=\log_2 \Big(2\pi e\big(\sigma^2+(1-\rho^2)(|g_{21}|^2+|g_{31}|^2)P_1\big)\Big),
\end{align}
so there must exist a constant $\alpha\in[0,1]$ such that
\begin{multline}
\label{alpha}
h(Y'|U,X_2)=\\
\log_2\Big( 2\pi e\big(\sigma^2+\alpha(1-\rho^2)(|g_{21}|^2+|g_{31}|^2)P_1\big)\Big).
\end{multline}
The term with $\alpha$ in the (\ref{G_d2d:outer:R3}) is obtained as follows:
\allowdisplaybreaks
\begin{align}
R_3&\le I(X_1;Y_2,Y_3|U,X_2)\notag \\
&\overset{(c)}= I(X_1;Y'|U,X_2)\notag\\
&=h(Y'|U,X_2)-h(Y'|U,X_1,X_2)\notag\\
&=h(Y'|U,X_2)-h\Bigg(\frac{g_{21}Z_2+g_{31}Z_3}{\sqrt{|g_{21}|^2+|g_{31}|^2}}\Bigg)\notag\\
&\overset{(d)}=
\mC\bigg(\frac{\alpha(1-\rho^2)(|g_{21}|^2+|g_{31}|^2)P_1}{\sigma^2}\bigg),
\end{align}
where step $(c)$ follows by Lemma \ref{lemma:Y'} and step $(d)$ follows by (\ref{alpha}).

Moreover, we derive that
\begin{align}
\label{h(Y2|UX2)}
h(Y_2|U,X_2) &= h(\omega Y'+Z'_2|U,X_2)\notag\\
&\overset{(e)}\ge \log_2\Big(2^{h(\omega Y'|U,X_2)}+2^{h(Z'_2|U,X_2)}\Big)\notag\\
&= \log_2\big(1+\alpha(1-\rho^2)|g_{21}|^2P_1\big),
\end{align}
where $\omega=\frac{g_{21}}{\sqrt{|g_{21}|^2+|g_{31}|^2}}$ and $Z'_2$ is previously defined in (\ref{Z'_2}); step $(e)$ follows by the entropy power inequality (EPI) since $\omega Y'\indep Z'_2$ given $(U,X_2)$. Consequently, we have
\begin{align}
R_1 &\le I(U;Y_2|X_2)\notag\\
&= h(Y_2|X_2) - h(Y_2|U,X_2)\notag\\
&\overset{(f)}\le \mC\bigg(\frac{(1-\rho^2)|g_{21}|^2P_1}{\sigma^2+\alpha(1-\rho^2)|g_{21}|^2P_1}\bigg),
\end{align}
where step $(f)$ is due to (\ref{h(Y2|UX2)}).

Next, we show the upper bounds on $R_1$ or $R_3$ that involve the auxiliary variable $V$. First, we derive the following chain of inequalities:
\begin{align}
I(X_1;Y_2,Y_3|V,X_2)&\overset{(g)}\le I(X_1;Y_2,Y_3|X_2)\notag\\
&=\mC\bigg(\frac{(1-\rho^2)(|g_{21}|^2+|g_{31}|^2)P_1}{\sigma^2}\bigg),
\end{align}
where step $(g)$ follows since $V\rightarrow X_1\rightarrow (Y_2,Y_3)$ form a Markov chain conditioned on $X_2$. Thus, we are guaranteed to find a constant $\beta\in[0,1]$ such that
\begin{align}
\label{beta}
R_1&\le I(X_1;Y_2,Y_3|V,X_2)\notag\\
&=\mC\bigg(\frac{\beta(1-\rho^2)(|g_{21}|^2+|g_{31}|^2)P_1}{\sigma^2}\bigg).
\end{align}
Finally, we can compute the second mutual information term in (\ref{DM_d2d:outer:cc}) as
\begin{align}
R_3 &\le I(V;Y_2,Y_3|X_2)\notag\\
&=I(X_1;Y_2,Y_3|X_2)-I(X_1;Y_2,Y_3|V,X_2)\notag\\
&\overset{(h)}=I(X_1;Y_2,Y_3|X_2)-\mC\bigg(\frac{\beta(1-\rho^2)(|g_{21}|^2+|g_{31}|^2)P_1}{\sigma^2}\bigg)\notag\\
&\le \mC\bigg(\frac{(1-\rho^2)(|g_{21}|^2+|g_{31}|^2)P_1}{\sigma^2+\beta(1-\rho^2)(|g_{21}|^2+|g_{31}|^2)P_1}\bigg),
\end{align}
where step $(h)$ is due to the identity in (\ref{beta}). We have established the set of inequalities (\ref{G_d2d:outer:R1}) and (\ref{G_d2d:outer:R3}). The verification of the remaining inequalities in (\ref{G_d2d:outer}) is similar to the proof of Proposition \ref{prop:G:outer}.

{
\section{Proof of Theorem \ref{theorem:G_d2d:gap}}
\label{appendix:gap}

This proof consists of two parts: we first show that the D2D rate splitting
method attains the capacity region to within 1 bit when $|g_{21}| < |g_{31}|$,
then show that the uplink rate splitting method achieves the same constant-gap
when $|g_{21}| \ge |g_{31}|$. For ease of notation, we use
$(\varrho_1,\varrho_2,\varrho_3,\varrho_4,\varrho_5)$ to denote each of the
terms on the right-hand sides of the inequalities in (\ref{G_d2d:outer:gap}),
which correspond to the upper bounds on $R_1$, $R_2$, $R_1+R_3$, $R_2+R_3$, and
$R_1+R_2+R_3$, respectively.

\subsection{Constant-Gap Optimality of D2D Rate Splitting}

We aim to show that the achievable rate region of Proposition
\ref{prop:G_d2d:inner1} is within 1 bit of the capacity region of the scalar
Gaussian FD cellular network with D2D when $|g_{21}| < |g_{31}|$. The main idea
is to show that the gap is less than or equal to 1 bit between the inner bound
(\ref{G_d2d:inner1}) and the outer bound (\ref{G_d2d:outer:gap}).

We begin with a special case where $|g_{21}|^2P_1\le\sigma^2$. Since the SNR of
the uplink channel from node 1 to node 2 is upper bounded by 1, removing this
channel (i.e., setting $g_{21}$ to zero) would cause at most 1 bit loss to each
of $(R_1,R_2,R_3)$. Actually, without the uplink channel, our channel model
reduces to a multiple access channel, for which the inner bound
(\ref{G_d2d:inner1}) coincides with the capacity region if we set $a=b=d=0$ and
$c=e=1$. Thus, (\ref{G_d2d:inner1}) must be within 1 bit of the capacity of the
original channel model.

We now assume that $|g_{21}|^2P_1>\sigma^2$. When $a=d=0$, $e=1$,
$c=\frac{\sigma^2}{|g_{21}|^2P_1}$, and $b=1-c$, the inner bound
(\ref{G_d2d:inner1}) is \allowdisplaybreaks
\begin{subequations}
\label{G_d2d:inner1:gap}
\begin{align}
R_1 &\le \mC\bigg(\frac{|g_{21}|^2P_1-\sigma^2}{2\sigma^2}\bigg),
    \label{G_d2d:inner1:gap:a}\\
R_2 &\le \mC\bigg(\frac{|g_{32}|^2P_2}{\sigma^2}\bigg),\\
R_1+R_3 &\le \mC\bigg(\frac{|g_{21}|^2P_1-\sigma^2}{2\sigma^2}\bigg)+
    \mC\bigg(\frac{|g_{31}|^2}{|g_{21}|^2}\bigg),\\
R_1+R_2+R_3 &\le \mC\bigg(\frac{|g_{21}|^2P_1-\sigma^2}{2\sigma^2}\bigg)\notag\\
&\quad+\mC\bigg(\frac{\sigma^2|g_{31}|^2/|g_{21}|^2+|g_{32}|^2P_2}{\sigma^2}\bigg),
    \label{G_d2d:inner1:gap:sum1}\\
R_1+R_2+R_3 &\le \mC\bigg(\frac{|g_{31}|^2P_1+|g_{32}|^2P_2}{\sigma^2}\bigg).
    \label{G_d2d:inner1:gap:e}
\end{align}
\end{subequations}
Use $(\eta_1,\eta_2,\eta_3,\eta_4,\eta_5)$ to denote respectively the terms on
the right-hand side of (\ref{G_d2d:inner1:gap:a})--(\ref{G_d2d:inner1:gap:e}).
We can bound the constant gap
$\delta$ between (\ref{G_d2d:inner1}) and (\ref{G_d2d:outer:gap}) by comparing
$(\eta_1,\eta_2,\eta_3,\eta_4,\eta_5)$ with
$(\varrho_1,\varrho_2,\varrho_3,\varrho_4,\varrho_5)$. Furthermore, it can be
shown that
\begin{subequations}
\label{G_d2d:inner1:gap_simple}
\begin{align}
\eta_1 &\ge \mC\bigg(\frac{|g_{21}|^2P_1}{\sigma^2}\bigg)-1,
    \label{G_d2d:inner1:gap_simple:R1}\\
\eta_2 &\ge \mC\bigg(\frac{|g_{32}|^2P_2}{\sigma^2}\bigg),
    \label{G_d2d:inner1:gap_simple:R2}\\
\eta_3 &\ge \mC\bigg(\frac{|g_{21}|^2P_1}{\sigma^2}\bigg)+
    \mC\bigg(\frac{|g_{31}|^2}{|g_{21}|^2}\bigg)-1,
    \label{G_d2d:inner1:gap_simple:R13}\\
\min\{\eta_4,\eta_5\} &\ge \mC\bigg(\frac{|g_{31}|^2P_1+|g_{32}|^2P_2}{\sigma^2}\bigg)-1.
    \label{G_d2d:inner1:gap_simple:R123}
\end{align}
\end{subequations}
Thus, with respect to $R_1+R_3$, the constant gap $\delta$ can be bounded by analyzing the difference between (\ref{G_d2d:inner1:gap_simple:R13}) and (\ref{G_d2d:outer:R1R3:gap}):
\begin{align}
2\delta &\le \varrho_3-\eta_3\notag\\
&\le \mC\bigg(\frac{(|g_{21}|^2+|g_{31}|^2)P_1}{\sigma^2}\bigg) - \mC\bigg(\frac{|g_{21}|^2P_1}{\sigma^2}\bigg)\notag\\
&\quad-\mC\bigg(\frac{|g_{31}|^2}{|g_{21}|^2}\bigg)+1\notag\\
&\overset{(a)}\le \mC\bigg(\frac{2|g_{31}|^2P_1}{\sigma^2}\bigg) - \mC\bigg(\frac{|g_{21}|^2P_1}{\sigma^2}\bigg)-\mC\bigg(\frac{|g_{31}|^2}{|g_{21}|^2}\bigg)+1\notag\\
&=\log_2\bigg(\frac{\big(|g_{21}|^2P_1\big)\big(\sigma^2+2|g_{31}|^2P_1\big)}{\big(\sigma^2+|g_{21}|^2P_1\big)\big(|g_{21}|^2+|g_{31}|^2\big)P_1}\bigg)+1\notag\\
&\overset{(b)}\le \log_2(2)+1\notag\\
&\le 2,
\end{align}
where step $(a)$ follows by $|g_{21}| < |g_{31}|$ and step $(b)$ follows by $|g_{21}|^2P_1>\sigma^2$. With respect to $R_1+R_2+R_3$, we have
\begin{align}
3\delta &\le \varrho_5-\min\{\eta_4,\eta_5\}\notag\\
&\le \mC\left(\frac{|g_{31}|^2P_1+|g_{32}|^2P_2+J}{\sigma^2}\right)+\mC\left(\frac{|g_{21}|^2P_1}{\sigma^2+|g_{31}|^2P_1}\right)\notag\\
&\quad-\mC\bigg(\frac{|g_{31}|^2P_1+|g_{32}|^2P_2}{\sigma^2}\bigg)+1\notag\\
&\overset{(c)}\le \mC\left(\frac{2(|g_{31}|^2P_1+|g_{32}|^2P_2)}{\sigma^2}\right)+\mC(1)\notag\\
&\quad-\mC\bigg(\frac{|g_{31}|^2P_1+|g_{32}|^2P_2}{\sigma^2}\bigg)+1\notag\\
&\le 3,
\end{align}
where step $(c)$ follows as $|g_{21}|\le|g_{31}|$.

Likewise, the constant gap can be established with respect to $R_1$ by showing that $\varrho_1-\mu_1\le1$, and with respect to $R_2$ by showing that $\varrho_2-\mu_2\le1$. Note that $\varrho_4$ is not used in this proof because there is no inner bound on $R_2+R_3$ in (\ref{G_d2d:inner1:gap}). Summarizing the above results gives a constant gap of 1 bit.

\subsection{Constant-Gap Optimality of Uplink Rate Splitting}

We further show that the achievable rate region of Proposition \ref{prop:G_d2d:inner2} is within 1 bit of the capacity region of the scalar Gaussian FD cellular network with D2D when $|g_{21}| \ge |g_{31}|$.



First consider a special where $|g_{31}|^2P_1\le\sigma^2$. Use
$(\nu_1,\nu_2,\nu_3,\nu_4)$ to denote respectively the terms on the right-hand
side of the inequalities in (\ref{G_d2d:inner2}) when $a=b=d=0$ and $c=e=1$,
which correspond to the inner bounds on $R_2$, $R_1+R_3$, $R_2+R_3$, and
$R_1+R_2+R_3$, respectively. Because $|g_{31}|^2P_1\le\sigma^2$, we have
\begin{subequations}
\label{G_d2d:inner2:small}
\begin{align}
\nu_1 &\ge \mC\bigg(\frac{|g_{32}|^2P_2}{2\sigma^2}\bigg),
    \label{G_d2d:inner2:small:R2}\\
\nu_2 &= \mC\bigg(\frac{|g_{21}|^2P_1}{\sigma^2}\bigg),
    \label{G_d2d:inner2:small:R13}\\
\nu_3 &\ge \mC\bigg(\frac{|g_{32}|^2P_2}{2\sigma^2}\bigg),
    \label{G_d2d:inner2:small:R23}\\
\nu_4 &\ge \mC\bigg(\frac{|g_{32}|^2P_2}{2\sigma^2}\bigg)+\mC\bigg(\frac{|g_{21}|^2P_1}{\sigma^2}\bigg).
    \label{G_d2d:inner2:small:R123}
\end{align}
\end{subequations}
It can be immediately seen that $\varrho_1-\nu_1\le1$ and thus $\delta\le1$ holds with respect to $R_1$. Further, we bound $\delta$ with respect to $R_1+R_3$ as follows:
\begin{align}
2\delta &\le \varrho_3-\nu_2\notag\\
&= \mC\bigg(\frac{(|g_{21}|^2+|g_{31}|^2)P_1}{\sigma^2}\bigg) -\mC\bigg(\frac{|g_{21}|^2P_1}{\sigma^2}\bigg)\notag\\
&\overset{(d)}\le \mC\bigg(\frac{|g_{21}|^2P_1+\sigma^2}{\sigma^2}\bigg) -\mC\bigg(\frac{|g_{21}|^2P_1}{\sigma^2}\bigg)\notag\\
&\le 1,
\end{align}
where step $(d)$ follows by $|g_{31}|^2P_1\le \sigma^2$. Next, we bound $\delta$ with respect to $R_2+R_3$:
\begin{align}
2\delta &\le \varrho_4-\nu_3\notag\\
&\le \mC\bigg(\frac{|g_{31}|^2P_1+|g_{32}|^2P_2+J}{\sigma^2}\bigg) -\mC\bigg(\frac{|g_{32}|^2P_2}{2\sigma^2}\bigg)\notag\\
&\le \mC\bigg(\frac{2\big(|g_{31}|^2P_1+|g_{32}|^2P_2\big)}{\sigma^2}\bigg) -\mC\bigg(\frac{|g_{32}|^2P_2}{2\sigma^2}\bigg)\notag\\
&\overset{(e)}<\mC\bigg(\frac{2\big(\sigma^2+|g_{32}|^2P_2\big)}{\sigma^2}\bigg) -\mC\bigg(\frac{|g_{32}|^2P_2}{2\sigma^2}\bigg)\notag\\
&\le 2,
\end{align}
where step $(e)$ is due to $|g_{31}|<|g_{21}|$. Finally, with respect to $R_1+R_2+R_3$, the constant gap $\delta$ is upper bounded as
\begin{align}
3\delta &\le \varrho_5-\nu_4\notag\\
&\le \mC\left(\frac{|g_{31}|^2P_1+|g_{32}|^2P_2+J}{\sigma^2}\right)+\mC\left(\frac{|g_{21}|^2P_1}{\sigma^2+|g_{31}|^2P_1}\right)\notag\\
&\quad - \mC\bigg(\frac{|g_{32}|^2P_2}{2\sigma^2}\bigg)-\mC\bigg(\frac{|g_{21}|^2P_1}{\sigma^2}\bigg)\notag\\
&\overset{(f)}\le \mC\left(\frac{2\big(|g_{31}|^2P_1+|g_{32}|^2P_2\big)}{\sigma^2}\right) - \mC\bigg(\frac{|g_{32}|^2P_2}{2\sigma^2}\bigg)\notag\\
&\le 2.
\end{align}
Thus, we verify that $\delta\le1$ when $|g_{31}|^2P_1 \le \sigma^2$.

The remainder of the proof assumes that $|g_{31}|^2P_1\ge\sigma^2$. We now set $a=d=0$, $c=\frac{\sigma^2}{|g_{31}|^2P_1}$, $e=1$, and $b=1-c$ in (\ref{G_d2d:inner2}), thus arriving at this inner bound:
\begin{subequations}
\label{G_d2d:inner2:gap}
\begin{align}
R_2 &\le \mC\bigg(\frac{|g_{32}|^2P_2}{2\sigma^2}\bigg),\\
R_1+R_3 &\le \mC\bigg(\frac{|g_{21}|^2P_1}{\sigma^2}\bigg),\\
R_2+R_3 &\le \mC\bigg(\frac{|g_{31}|^2P_1+|g_{32}|^2P_2-\sigma^2}{2\sigma^2}\bigg),\\
R_1+R_2+R_3 &\le \mC\bigg(\frac{|g_{31}|^2P_1+|g_{32}|^2P_2-\sigma^2}{2\sigma^2}\bigg)\notag\\ &\quad+\mC\bigg(\frac{|g_{21}|^2}{|g_{31}|^2}\bigg).
    \label{G_d2d:inner2:gap:R123}
\end{align}
\end{subequations}
Let $(\varphi_1,\varphi_2,\varphi_3,\varphi_4)$ be respectively the terms on the right-hand side of the above inequalities.
With respect to $R_1+R_2+R_3$, we bound the constant gap $\delta$ as follows:
\begin{align}
3\delta &\le \varrho_5-\varphi_4\notag\\
&\le \mC\left(\frac{2\big(|g_{31}|^2P_1+|g_{32}|^2P_2\big)}{\sigma^2}\right)+\mC\left(\frac{|g_{21}|^2P_1}{\sigma^2+|g_{31}|^2P_1}\right)\notag\\
&\quad-\mC\bigg(\frac{|g_{31}|^2P_1+|g_{32}|^2P_2-\sigma^2}{2\sigma^2}\bigg)-\mC\bigg(\frac{|g_{21}|^2}{|g_{31}|^2}\bigg)\notag\\
&= \mC\left(\frac{2\big(|g_{31}|^2P_1+|g_{32}|^2P_2\big)}{\sigma^2}\right)+\mC\left(\frac{|g_{21}|^2P_1}{\sigma^2+|g_{31}|^2P_1}\right)\notag\\
&\quad-\mC\bigg(\frac{|g_{31}|^2P_1+|g_{32}|^2P_2}{\sigma^2}\bigg)-\mC\bigg(\frac{|g_{21}|^2}{|g_{31}|^2}\bigg)+1\notag\\
&\le 2.
\end{align}
It can be also shown that $\delta\le\varrho_2-\varphi_1\le1$, $2\delta\le\varrho_3-\varphi_2\le2$, and $2\delta\le\varrho_4-\varphi_3\le2$. The proof is then complete.
}

\bibliographystyle{IEEEbib}
\bibliography{IEEEabrv,strings}

\begin{IEEEbiographynophoto}{Kaiming Shen}
(S'13-M'20) received the B.Eng. degree in information security and the B.S. degree in mathematics from Shanghai Jiao Tong University, Shanghai, China in 2011, then the M.A.Sc. and Ph.D. degrees in electrical and computer engineering from the University of Toronto, Ontario, Canada in 2013 and 2020, respectively.

Since 2020, he has been an Assistant Professor with the School of Science and Engineering at the Chinese University of Hong Kong (Shenzhen), China. His main research interests include optimization, wireless communications, data science, and information theory.
\end{IEEEbiographynophoto}

\newpage

\begin{IEEEbiographynophoto}{Reza K. Farsani}
was born in Farsan, Iran, in 1986. He received the double
major B.Sc. degrees in electrical engineering and pure mathematics from
Sharif University of Technology, Tehran, Iran, in 2010, and the M.S. degree in
electrical engineering from the University of Waterloo, Waterloo, ON, Canada,
in 2016. He is currently pursuing the Ph.D. degree at
the University of Toronto, Toronto, Canada.

From 2010 to 2014, he was a Research Scientist with the Institute for
Research in Fundamental Sciences (IPM), Tehran, where he accomplished a
research project on fundamental limits of communications in interference networks. From 2016 to 2017, he was a Research Assistant with the Department
of Electrical and Computer Engineering, University of Waterloo. His research
interests include information and communication theory, machine learning,
and statistics.

Mr. Farsani received the Silver Medal from the Iranian Students Mathematical Olympiad in 2003.
\end{IEEEbiographynophoto}

\begin{IEEEbiographynophoto}{Wei Yu}
(S'97-M'02-SM'08-F'14) received the B.A.Sc. degree in Computer
Engineering and Mathematics from the University of Waterloo, Waterloo,
Ontario, Canada in 1997 and M.S. and Ph.D. degrees in Electrical
Engineering from Stanford University, Stanford, CA, in 1998 and 2002,
respectively. Since 2002, he has been with the Electrical and Computer
Engineering Department at the University of Toronto, Toronto, Ontario,
Canada, where he is now Professor and holds a Canada Research Chair
(Tier 1) in Information Theory and Wireless Communications. 

Prof. Wei Yu
serves as the President of the IEEE Information Theory Society in 2021,
and has served on its Board of Governors since 2015. He served as
the Chair of the Signal Processing for Communications and Networking
Technical Committee of the IEEE Signal Processing Society in 2017-18.
Prof. Wei Yu was an IEEE Communications Society Distinguished Lecturer
in 2015-16. He is currently an Area Editor for the IEEE Transactions on
Wireless Communications, and in the past served as an Associate Editor
for IEEE Transactions on Information Theory, IEEE Transactions on
Communications, and IEEE Transactions on Wireless Communications. 

Prof.
Wei Yu is a Fellow of the Canadian Academy of Engineering, and a member
of the College of New Scholars, Artists and Scientists of the Royal
Society of Canada. He received the Steacie Memorial Fellowship in 2015,
the IEEE Marconi Prize Paper Award in Wireless Communications in 2019,
the IEEE Communications Society Award for Advances in Communication in
2019, the IEEE Signal Processing Society Best Paper Award in 2017 and
2008, the Journal of Communications and Networks Best Paper Award in
2017, and the IEEE Communications Society Best Tutorial Paper Award in
2015.
\end{IEEEbiographynophoto}


\end{document}